\documentclass[12pt,preprint]{aastex}

\slugcomment{WU-AP/258/06}

\shorttitle{Stellar Core Collapse and Neutrino Emission}
\shortauthors{Nakazato, Sumiyoshi, \& Yamada}

\begin{document}


\title{Numerical Study on Stellar Core Collapse and Neutrino Emission: \\
     Probe into the Spherically Symmetric Black Hole Progenitors with 3
     - 30$M_\odot$ Iron Cores}


\author{Ken'ichiro Nakazato\altaffilmark{1}, Kohsuke
Sumiyoshi\altaffilmark{2,3} and Shoichi Yamada\altaffilmark{1,4}}

\email{nakazato@heap.phys.waseda.ac.jp}


\altaffiltext{1}{Department of Physics,
Waseda University, 3-4-1 Okubo, Shinjuku, Tokyo 169-8555, Japan}
\altaffiltext{2}{Numazu College of Technology, Ooka 3600, Numazu,
Shizuoka 410-8501, Japan}
\altaffiltext{3}{Division of Theoretical Astronomy, National
Astronomical Observatory of Japan, 2-21-1 Osawa, Mitaka, Tokyo 181-8588,
Japan}
\altaffiltext{4}{Advanced Research Institute for Science \& Engineering, 
Waseda University, 3-4-1 Okubo, Shinjuku, Tokyo 169-8555, Japan}


\begin{abstract}

 The existence of various anomalous stars, such as the first stars in
 the universe or stars produced by stellar mergers, has been recently
 proposed. Some of these stars will result in black hole formation. In
 this study, we investigate iron core collapse and black hole formation
 systematically for the iron-core mass range of 3 - 30$M_\odot$, which
 has not been studied well so far. Models used here are mostly
 isentropic iron cores that may be produced in merged stars in the
 present universe but we also employ a model that is meant for a
 Population III star and is obtained by evolutionary calculation. We
 solve numerically the general relativistic hydrodynamics and neutrino
 transfer equations simultaneously, treating neutrino reactions in
 detail under spherical symmetry. As a result, we find that massive iron
 cores with $\sim10M_\odot$ unexpectedly produce a bounce owing to the
 thermal pressure of nucleons before black hole formation. The features
 of neutrino signals emitted from such massive iron cores differ in time
 evolution and spectrum from those of ordinary supernovae. Firstly, the
 neutronization burst is less remarkable or disappears completely for
 more massive models because the density is lower at the
 bounce. Secondly, the spectra of neutrinos, except the electron type,
 are softer owing to the electron-positron pair creation before the
 bounce. We also study the effects of the initial density profile,
 finding that the larger the initial density gradient is, the more
 steeply the neutronization burst declines. Further more, we suggest a
 way to probe into the black hole progenitors from the neutrino emission
 and estimate the event number for the currently operating neutrino
 detectors.

\end{abstract}





\section{Introduction} \label{intro}

Various anomalous stars, such as the first stars in the universe
(so-called Population III stars) or stars produced by stellar mergers in
stellar clusters, are being studied recently. As for the Population III
stars, it is suggested theoretically that they are much more massive
($M\gtrsim100M_{\odot}$) than stars of later generations (e.g., Nakamura
\& Umemura 2001). On the other hand, $N$-body simulations show that the
runaway mergers of massive stars occur and that very massive
($M\gtrsim100M_{\odot}$) stars are formed in a young compact stellar
cluster (e.g., Portegies Zwart 1999). Especially notable is a newly
suggested formation scenario for supermassive black holes which requires
the formation of intermediate-mass black holes by the collapse of merged
stars in very compact stellar clusters (e.g., Ebisuzaki et al. 2001). If
these anomalous stars collapse to black holes without supernova
explosions, it is supposedly difficult to be hard to probe into their
progenitors. One possible means of such a probe is, we think, to examine
the neutrinos emitted during the black hole formation. For this purpose,
systematic studies on black hole formation, including effects of the
neutrinos, are needed.

So far, various numerical simulations of supernova explosions have been
done by many authors. Similarly, numerical studies on black hole
formation have also been recently produced (e.g., Fryer 1999; Linke et
al. 2001; Fryer et al. 2001, hereafter FWH01; Sekiguchi \& Shibata 2005,
hereafter SS05; Nakazato et al. 2006, hereafter NSY06; Sumiyoshi et
al. 2006, hereafter SYSC06). Fryer (1999) classified core collapse into
three types. a) Stars with $M\lesssim25M_{\odot}$ make explosions and
produce neutron stars. b) Stars ranging
$25M_{\odot}\lesssim~M\lesssim40M_{\odot}$ also result in explosions but
produce black holes via fallback. c) For $M\gtrsim40M_{\odot}$, the
shock produced at the bounce can neither propagate out of the core nor
make explosions. In any case, the core bounces once. SYSC06 computed
fully general relativistic hydrodynamics under spherical symmetry,
taking into account the reactions and transports of neutrinos in detail
and confirmed class c) for the collapse of a progenitor with
$40M_{\odot}$. On the other hand, much more massive stars result in
black hole formation without bounce. SS05 studied the criterion for the
collapse without bounce. Their computations are fully general
relativistic, and they investigated the dynamics systematically, varying
the initial mass and rotation. They concluded that non-rotating iron
cores with a mass of $M_\mathrm{iron}\gtrsim2.2M_{\odot}$ collapse to
black holes without bounce. However, they employed the phenomenological
equation of state and did not consider the effects of neutrinos.

There are studies also on a collapse of very massive stars in a context
of the evolution of Population III stars. As mentioned already,
Population III stars may be very massive, $M\gtrsim100M_{\odot}$. It is
supposed that the pair creation of electrons and positrons makes a star
unstable during the helium burning phase if they do not lose much of
their mass during the quasi-static evolutions because of zero 
metallicity. Stars with $\lesssim260M_\odot$ reverse the collapse by
rapid nuclear burnings and explode to pieces, which are called 
pair-instability supernovae, while more massive stars cannot halt the
collapse and form black holes (e.g., Heger et al. 2003). Note that,
however, these numbers are still uncertain at present (e.g., Ohkubo et
al. 2006). Assuming that Population III stars with
$M\gtrsim300M_{\odot}$ are formed and evolve without mass loss, FWH01
and NSY06 showed that they collapse without bounce for spherically 
symmetric models under fully general relativistic computations while
NSY06 treated the neutrino transport more in detail than FWH01. FWH01
also computed the collapse of a rotating star with $300M_{\odot}$ under
Newtonian gravity and showed that it has a weak bounce and then
recollapses to a black hole immediately. As for the collapse of
supermassive stars with $M\gtrsim5\times10^5M_{\odot}$, Linke et
al. (2001) found that they form black holes without bounce before
becoming opaque to neutrinos.

The black hole formation of stars in the mass range between
$\sim 100M_{\odot}$ and $\sim 260M_{\odot}$, which corresponds to the
iron-core mass range between $\sim 3M_{\odot}$ and $\sim 30M_{\odot}$,
has not been studied well so far. This is because they are supposed to
explode as pair-instability supernovae during the quasi-static
evolutions if they are single stars. Recently, on the other hand, stars
produced by stellar mergers in a young compact stellar cluster were
studied in detail and their evolutionary paths are beginning to be
revealed (Suzuki et al. 2007). While they do not calculate the
evolutions of these stars up to the black hole formation, we speculate,
as in \S~\ref{prog}, that they may avoid the explosions as
pair-instability supernovae and form a massive iron core of the
above-mentioned range. Therefore, we investigate, in this study, the
iron core collapses systematically for the iron core masses of
$3M_{\odot}$ and $30M_{\odot}$, although there is no evidence to show
their existence so far.

To be more specific, we assume that the mass of an iron core is mainly
determined by the entropy per baryon, and our investigation is done
systematically for entropy. We solve the general relativistic
hydrodynamics under spherical symmetry. We also solve the neutrino
transfer equations simultaneously, treating neutrino reactions in
detail. In addition to the isentropic iron core models, we employ the
realistic stellar model of $100M_{\odot}$ and zero metallicity,
supposedly a Population III star, by Nomoto et al. (2005). We address
the issues concerning the black hole formation of the merged stars in
connection with our study and estimate the neutrino event number for the
currently operating detectors. We also suggest a way to probe into the
progenitors from the detection. We hope that this study will be not only
a reference for future multi-dimensional computations but also provide a
basis for neutrino astrophysics in the black hole formation.

\section{Initial Models and Numerical Methods} \label{mdlmthds}

At first, we construct the iron core models, which will later be used as
initial models for the dynamical simulation of the collapse. The
progenitor with $40M_{\odot}$ in SYSC06 has an entropy per baryon,
$s\sim1.5k_B$ and an iron core mass, $M_\mathrm{iron}=1.98M_{\odot}$,
whereas the massive Population III star models in NSY06 have $s>16k_B$
and $M_\mathrm{iron}>50M_{\odot}$. In this study, we intend to bridge
the gap of the black hole progenitors and discuss the neutrino emission
systematically for this range. Unfortunately, realistic models of the
progenitors for this range are rare, and ``systematic'' models for them
are absent up to the present, since their astrophysical counterparts are
not well known, as mentioned already. Therefore, we construct the
initial models by ourselves.

\begin{figure}
\plotone{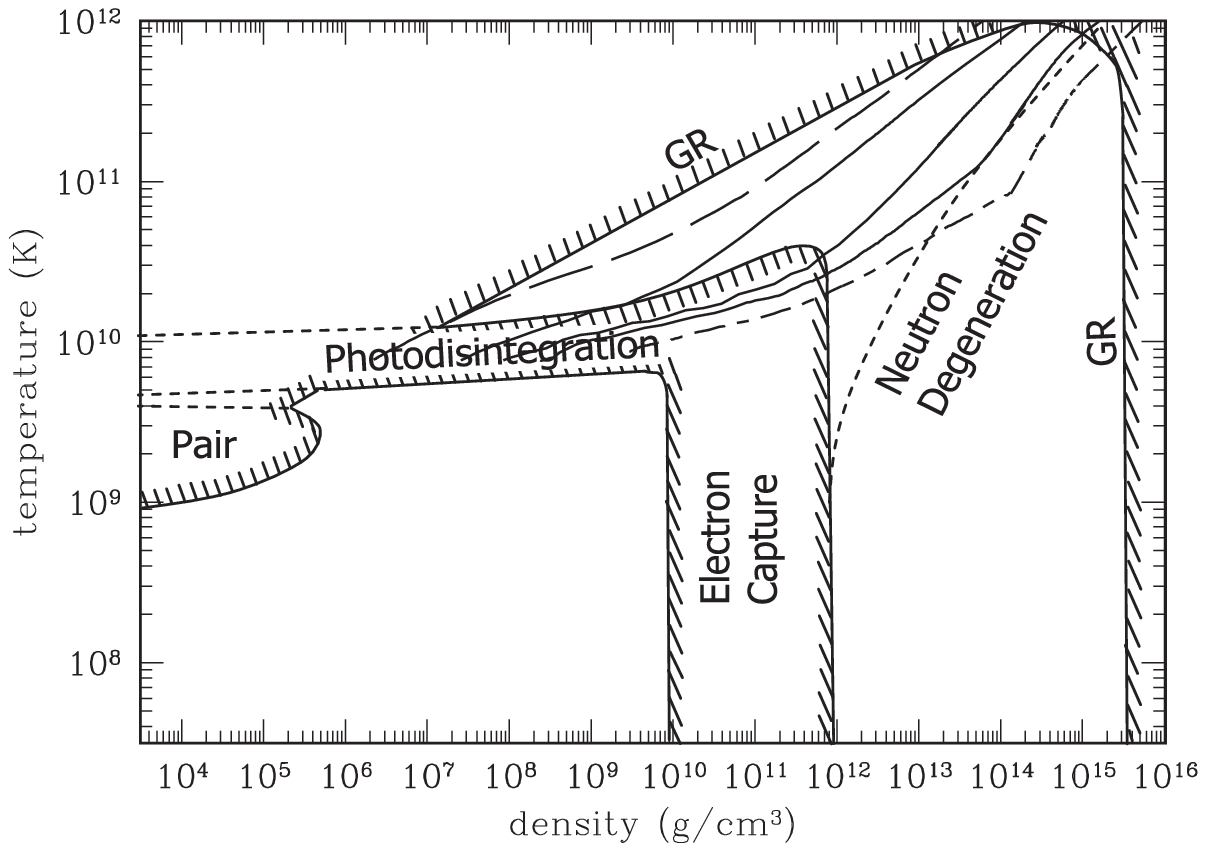}
\caption{The evolution of the central density and temperature for
 various models which result in black hole formation. The dot-dashed
 line is for a realistic progenitor with the initial mass $40M_\odot$
 in SYSC06 and the long-dashed line is for a Population III star with the
 initial mass $10500M_\odot$ ($s=74.75k_\mathrm{B}$) in NSY06. Three
 solid lines are for  our models in this paper, and each line corresponds
 to model~1a ($s=3k_\mathrm{B}$), 3a ($s=5k_\mathrm{B}$) and 5a
 ($s=10k_\mathrm{B}$), from right to left. The shaded area represents
 a gravitationally unstable region by labeled physical processes.}
\label{result}
\end{figure}

We assume that the iron cores in equilibrium configurations collapse by
photodisintegration, as is the case for the onset of ordinary core
collapse supernovae. We obtain the initial models, solving the
Oppenheimer-Volkoff equation with the equation of state by Shen et
al. (1998a, 1998b) assuming isentropy and the electron fraction
$Y_e=0.5$ throughout the core. We define the mass of the iron core,
$M_\mathrm{iron}$, as the mass coordinate where the temperature is
$5\times 10^9$~K, whereas we set the outer boundary at a much larger
radius so as not to affect the dynamics. For the systematic analysis, we
set the initial central temperature as
$T_\mathrm{initial}=7.75\times10^9$~K, which is slightly higher than the
critical temperature for the photodisintegration (Figure~\ref{result}),
and generate models~1a-6a with the values of entropy per baryon,
$s=3k_B$-$13k_B$, which have not been studied well so far, as mentioned
above. In order to investigate the ambiguity in the onset of collapse,
we also adopt a model (model~2b) with the same initial entropy per
baryon as model~2a ($s=4k_B$) but having half the central density. The
key parameters of these models are summarized in
Table~\ref{bounce_result}. In addition, we also employ the realistic 
stellar model of $100M_{\odot}$ with a vanishing metallicity by Nomoto
et al. (2005) in order to validate the isentropic models. This model is
supposedly a Population III star and resides in the range
$s=3k_B$-$13k_B$. 

As a next step, we compute the dynamics of spherically symmetric
gravitational collapse with the neutrino transport. As for our numerical
methods, we follow NSY06 and use the general relativistic implicit
Lagrangian hydrodynamics code, which solves simultaneously the neutrino
Boltzmann equations (Yamada 1997 ; Yamada et al. 1999 ; Sumiyoshi et
al. 2005). We consider four species of neutrino, $\nu_e$, $\bar\nu_e$,
$\nu_\mu$ and $\bar\nu_\mu$, assuming that $\nu_\tau$ and $\bar\nu_\tau$
are the same as $\nu_\mu$ and $\bar\nu_\mu$, respectively, and take into
account 9 neutrino reactions listed in NSY06. We use 127 radial mesh
points, while 12 and 4 mesh points are used for energy- and angular
distribution of neutrino, respectively. In order to assess the
convergence of our results, we compute models with higher
resolutions. They have the same initial conditions as model~2a. For
model~2m, the number of radial mesh points is increased to 255. Model~2e
use 18 mesh points for the energy spectrum while model~2g has 6 mesh
points for the angular distribution.

It is noted that our method allows us to follow the dynamics with no
difficulty up to the apparent horizon formation.  The existence of the
apparent horizon is the sufficient condition for the formation of a
black hole (or, equivalently, of an event horizon). For the Misner-Sharp
metric (Misner \& Sharp 1964) adopted in our computations, the radius of
the apparent horizon is written as
\begin{equation}
r = \frac{2G\widetilde m}{c^2},
\label{ah}
\end{equation}
where $c$ and $G$ are the velocity of light and the gravitational
constant, respectively (van Riper 1979). $r$ is the circumference radius
and $\widetilde m$ is the gravitational mass inside $r$. Since our
models are spherically symmetric, there is no difficulty in finding the
horizon.

\section{Results and Discussions} 

In this section, we show the results of our computations and discuss
them. We study the dynamics of the collapse in \S~\ref{bm} and
investigate the features of the neutrinos emitted during the collapse in
\S~\ref{ns}. In \S~\ref{ivd}, we investigate the role of the initial
velocity or the deviation from equilibrium. We also make a comparison
with the realistic progenitor models in \S~\ref{rsm}. Finally, we
mention about the astronomical counterparts of our models and the
possibility of the probe into the progenitors of the events in
\S~\ref{ai}.

To overview the characteristics of the models which we surveyed, we show
in Figure~\ref{result}, the evolution of the central density and
temperature of our results together with those of other simulations of
black hole formation. The trajectories of the current models shown by
solid lines are between those of previous models reflecting the
different values of entropy. From this figure, we can recognize that our
investigation bridges the gap between two previous studies, SYSC06 and
NSY06.

\subsection{Dynamical Features}\label{bm}

\begin{figure}
\plotone{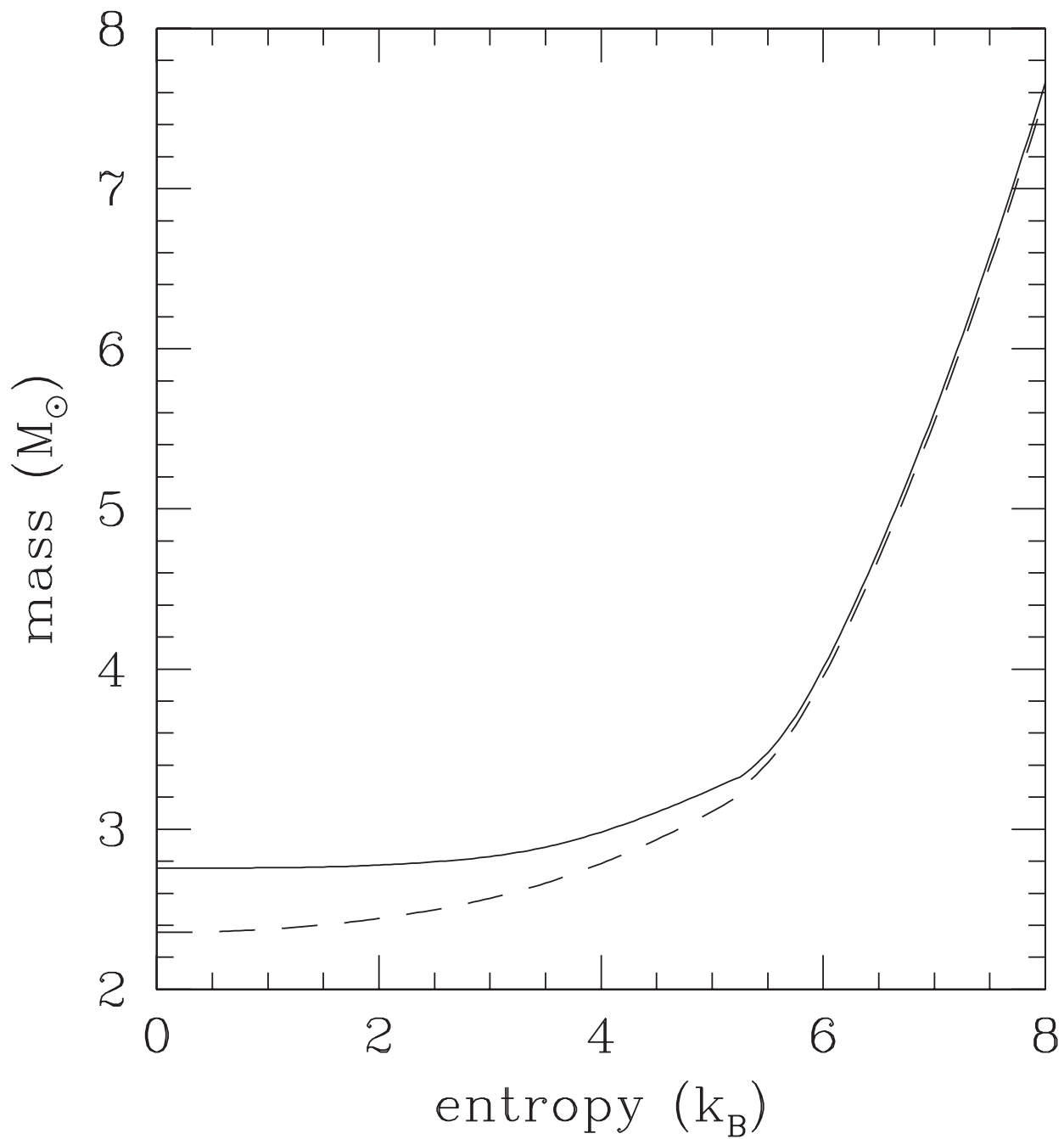} 
\caption{The maximum mass of the neutron star assuming isentropy and a
 constant electron fraction which is isentropic ($Y_e=0.1$) under the
 equation of state by Shen et al. (1998a, 1998b). The solid and dashed
 lines represent the maximum mass in the sense of the baryon rest mass
 and the gravitational mass, respectively.}
\label{maxmasf} 
\end{figure}

It is known that ordinary supernovae with $s\sim1k_\mathrm{B}$ bounce
because their central density exceeds the nuclear density
($\sim2.5\times10^{14}\mathrm{g\,cm^{-3}}$) and pressure drastically
increases. From our computations, we find that models with
$3k_\mathrm{B}\leq s\leq7.5k_\mathrm{B}$ ($M\leq10.6M_\odot$) have a
bounce and that they recollapse to black holes. On the other hand,
models with $s>7.5k_\mathrm{B}$ ($M>10.6M_\odot$) collapse to black
holes directly without bounce. We show the evolution of core collapse in
Figure~\ref{maxmasf} for two representative cases.

\begin{figure}
\plotone{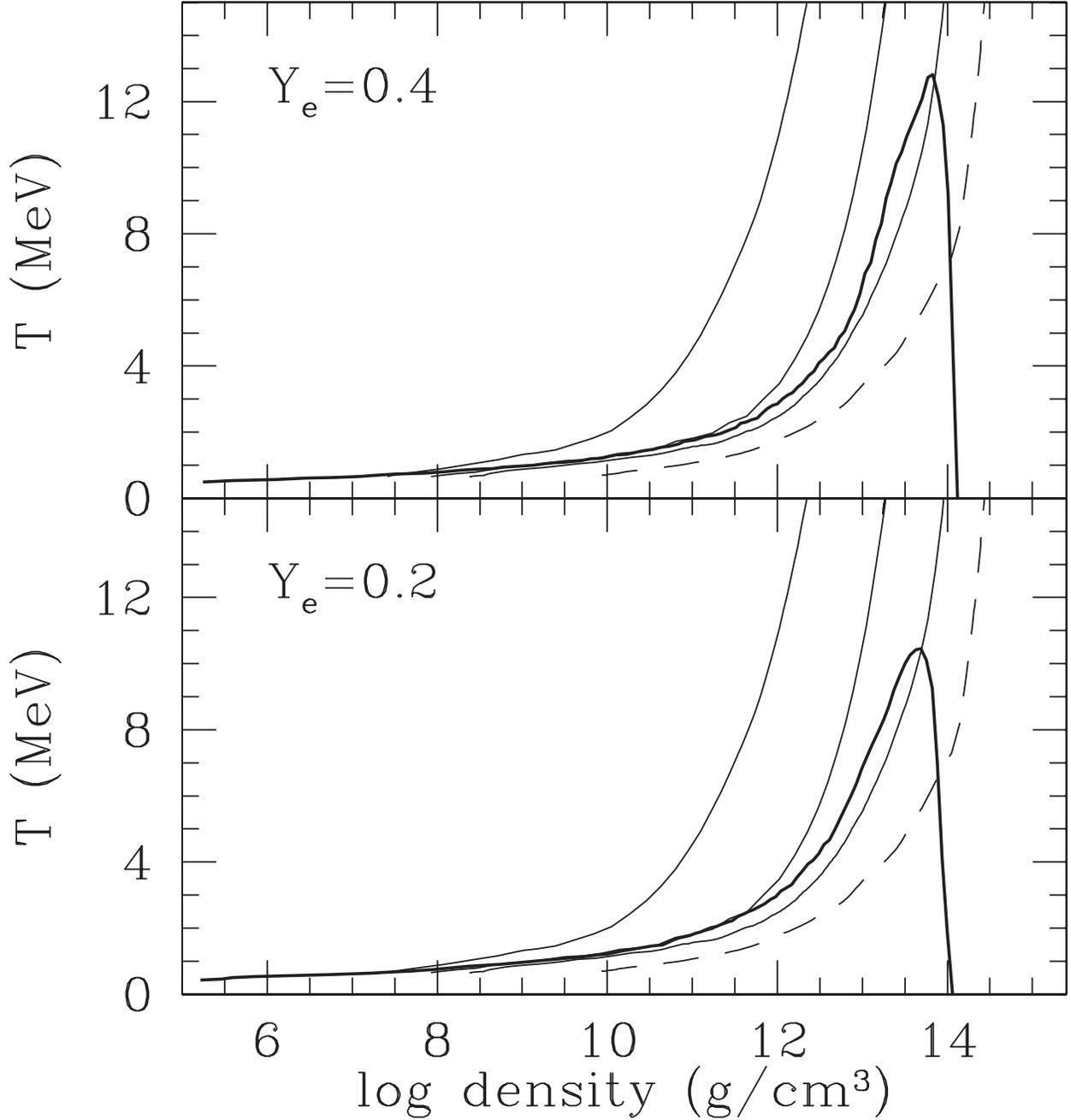} 
\caption{Phase diagram in $\rho-T$ plane from Shen et al. (1998b) for
 fixed electron fraction, $Y_e$ (thick lines). The nucleus exists in the
 region below these thick lines. The phase boundaries depend on $Y_e$,
 whereas the same trajectories are plotted for the upper panel and the
 lower panel. The dashed line represents the evolution of the central
 density and temperature for the ordinary supernova progenitor with the
 initial mass $15M_\odot$ (Sumiyoshi et al. 2005), and the solid lines
 do the same for the progenitors studied. Each line corresponds to
 models~1a ($s=3k_\mathrm{B}$), 3a ($s=5k_\mathrm{B}$) and 5a
 ($s=10k_\mathrm{B}$), from right to left.}
\label{phase}
\end{figure}

In the case of $3k_\mathrm{B}\leq s\leq7.5k_\mathrm{B}$, it is noted
that the bounce mechanism of the core with $s\geq3k_\mathrm{B}$ is not
the same as that of ordinary supernovae. The high entropy cores bounce
because of the thermal pressure of nucleons at sub-nuclear density. We
can see this fact from the evolutions of central density and temperature
in the phase diagram of the nuclear matter at $Y_e=0.4$ and $0.2$
(Figure~\ref{phase}). We note that for all models at the center,
$Y_e\sim0.4$ and $Y_e\sim0.2$ when $T\sim1$~MeV and $T\sim10$~MeV,
respectively. These figures show that the models with higher entropies
go from the non-uniform mixed phase of nuclei and free nucleons to the
classical ideal gas phase of thermal nucleons and $\alpha$ particles,
whereas that of an ordinary supernova goes into the uniform nuclear
matter phase. In the ideal gas phase, the number of non-relativistic
nucleons and $\alpha$ particles is comparable to that of relativistic
electrons. Since the adiabatic index of non-relativistic gas is
$\gamma=\frac{5}{3}$ and that of relativistic gas is
$\gamma=\frac{4}{3}$, the collapse is halted and bounce occurs.

\begin{figure}[h]
\begin{center}
\plotone{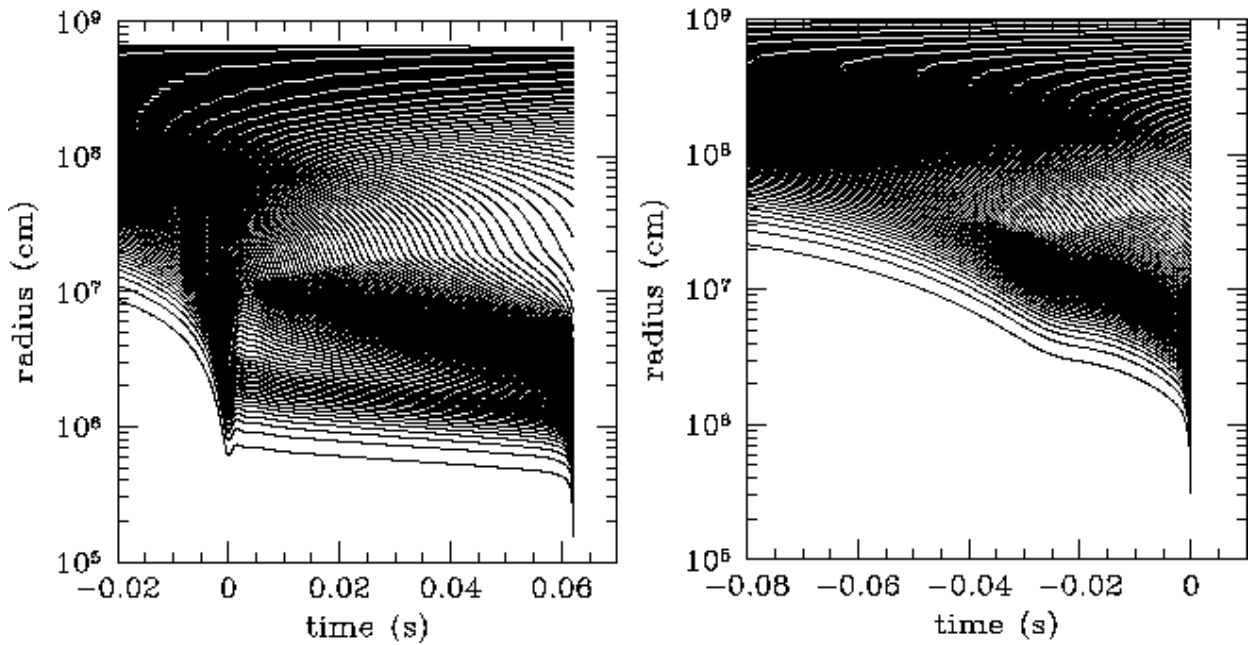}
\caption{Radial trajectories of mass elements. The left panel is for
 model~2a ($s=4k_B$); time is measured from the bounce. The right panel is
 for model~5a ($s=10k_B$); time is measured from the point at which the
 apparent horizon is formed.}
\label{shell4} 
\end{center}
\end{figure}

Because this bounce is weak and the shock is stalled, the inner core (or
the protoneutron star) grows beyond the maximum mass of the neutron star
and recollapses to a black hole soon (left panel of
Figure~\ref{shell4}). In Figure~\ref{maxmasf}, we show the maximum mass
of the neutron star assuming isentropy and the constant electron
fraction ($Y_e=0.1$) under the equation of state by Shen et
al. (1998a, 1998b). It is noted that the maximum mass is larger than
$3M_\odot$ for the neutron star with high entropies,
$s\gtrsim4k_\mathrm{B}$. Since the maximum mass of the neutron star
depends on the equation of state, it should be remind that the time
interval from the bounce to the recollapse also depends on it
(SYSC06). We will refer to this point again later.

In Table~\ref{bounce_result}, we show the inner core mass, central
density, temperature and adiabatic index at the bounce together with the
interval time from the bounce to the apparent horizon formation. We can
recognize that the density and the adiabatic index at the bounce get
lower for the models with higher initial entropies. These features
indicate that the bounce is not due to the nuclear force but to the
thermal pressure of non-relativistic gas for high entropy
cores. Moreover, the interval time from the bounce to the apparent
horizon formation is shorter for the higher entropy cores. This is
because the initial mass of the iron core ($M_\mathrm{iron}$) is larger
than the maximum mass of the neutron star ($M_\mathrm{max}$) for the
models with high entropies and they can collapse to black holes quickly.

We also show the results for the models with higher resolutions in
Table~\ref{bounce_result}. The central density and the adiabatic index
at bounce, which are key parameters in our analysis, are not very
different for models~2a, 2g and 2m. The central density at bounce of
model~2e, which has 1.5 times finer energy mesh, is different by 14\%
from that of model~2a. This is because neutrinos affect the entropy
variations before the neutrino trapping. In fact, the central entropy at
bounce of model~2a is $3.50k_\mathrm{B}$ while that of model~2e is
$3.62k_\mathrm{B}$. However, qualitative features of their bounces are
not changed. On the other hand, the interval times from the bounce to
the apparent horizon formation are different by $\lesssim$ 15\% for
models~2a, 2e, 2g and 2m. This is because the start point of the
recollpase is roughly determined by the maximum mass of the neutron star
as mentioned already. Since the mass accretion rate is of the order of
$10M_{\odot}\,\mathrm{s}^{-1}$ during this phase in our models, the
difference of $0.1M_{\odot}$ in the maximum mass means the difference of
10ms in the interval time, which is close to the discrepancies found
here. Thus, the precise determination of the interval time is difficult
in general. However, its dependence on the initial entropy is well
established.

We compare our results with other studies. In SS05, the non-rotating
models with an iron core mass of $\gtrsim2.28M_{\odot}$ end up with
black holes without bounce. In our models, on the other hand, it is
shown that the iron core with $\lesssim10.6M_{\odot}$ (or the initial
entropy $s\lesssim7.5k_\mathrm{B}$) has bounce before black hole
formation. This discrepancy comes from the fact that their equation of
state is parametric and does not take into account properly the effects
of thermal nucleons in the collapsing phase. On the other hand, the
rotating Population III star with $\sim300M_{\odot}$ has a weak bounce
at $\rho_c\sim10^{12}\mathrm{g\,cm^{-3}}$ in FWH01, and these authors
adopt a realistic equation of state (Herant et al. 1994). Since the
rotation tends to produce a bounce, we can predict that the bounce is
inevitable for an iron core with the mass $\lesssim10M_{\odot}$
irrespective of rotations, and the effects of thermal nucleons are
crucial.

In the high entropy case $s>7.5k_\mathrm{B}$, more massive cores do not
have a bounce but form an accretion shock before the apparent horizon
formation. This is because, the outer region keeps collapsing
supersonically while the central region becomes gravitationally stable
by the thermal pressure of non-relativistic gas. We can see this feature
in the right panel of Figure~\ref{shell4}. As the initial mass gets
larger, the transition occurs smoothly from the collapse with bounce to
the one without bounce. Incidentally, the features of direct collapse
are almost the same as those for the Population III models in NSY06,
where a detailed analysis can be found.

\subsection{Neutrino Signals}\label{ns}

In this section, we discuss neutrino emission during core collapse. As
mentioned already, we compute the collapse until the formation of the
apparent horizon. However, the location of the event horizon is not
known for our models although it is proved mathematically that the event
horizon is always located outside the apparent horizon. Moreover, the
numerical difficulty prevents us from computing the dynamics until the
apparent horizon swallows the shock surface entirely. Because of these
facts, the total energy and number of emitted neutrinos have some
ambiguities. In this study, we estimate the upper and lower limits for
the total energy and number of emitted neutrinos, following NSY06. The
upper limit is obtained with an assumption that all neutrinos in the
region between the shock surface and the neutrino sphere flow out
without being absorbed or scattered. For the lower limit, on the other
hand, we assume that all neutrinos in this region are trapped and do not
come out. Fortunately, for the models with bounce, these ambiguities are
minor, compared with the direct collapse models in NSY06, since the
duration from the shock formation to the apparent horizon formation is
longer and almost all neutrinos are emitted during this phase.

The calculated results of the neutrino emission are summarized in 
Table~\ref{nutr-eg}. It is noted that we assume that $\nu_\tau$
($\bar\nu_\tau$) is the same as $\nu_\mu$ ($\bar\nu_\mu$), and that the
luminosities of $\nu_\mu$ and $\bar\nu_\mu$ are almost identical because
they have the same reactions and because the difference of coupling
constants is minor. In the following, ignoring this tiny difference, we 
denote these four species as $\nu_x$ collectively. In
Table~\ref{nutr-eg}, we can recognize that the total energy does not
change monotonically with the initial entropy of the core. This is
because the duration of the neutrino emission is longer for the lower
entropy models, while the duration is shorter and the neutrino
luminosity is larger for the higher entropy models.

\begin{figure} 
\plotone{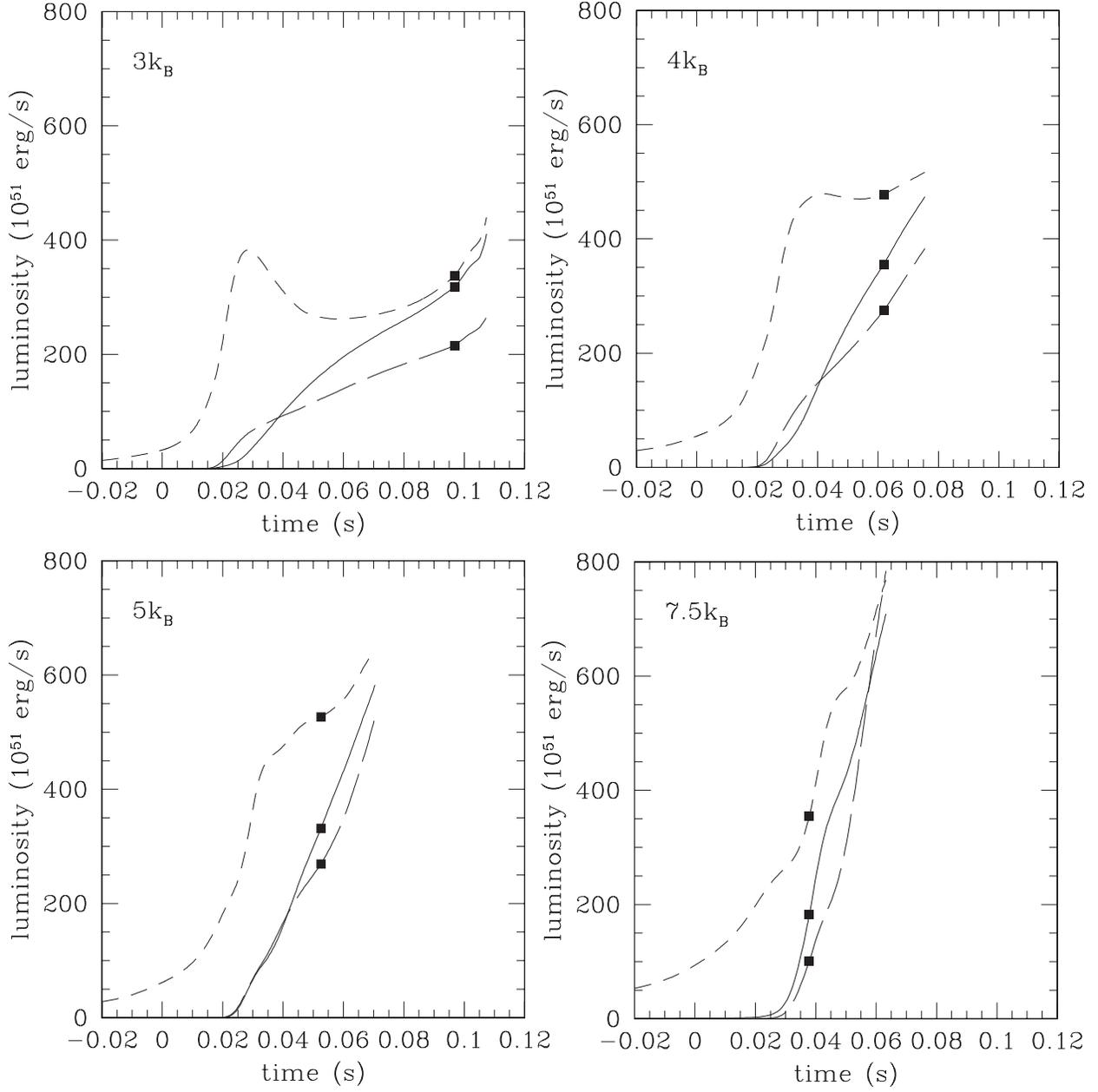} 
\caption{Luminosities of $\nu_e$ (short-dashed line), $\bar \nu_e$
 (solid line) and $\nu_x$ (long-dashed line) as a function of $t$, where
 $\nu_x$ stands for $\mu$- and $\tau$-neutrinos and their
 anti-particles. Squares show the time when the apparent horizon is
 formed. Upper left, upper right, lower left and lower right panels are
 for models~1a ($s=3k_\mathrm{B}$), 2a ($s=4k_\mathrm{B}$), 3a
 ($s=5k_\mathrm{B}$) and 4a ($s=7.5k_\mathrm{B}$), respectively.} 
\label{lumi} 
\end{figure}

In Figure~\ref{lumi}, we show the time evolutions of neutrino luminosity
for several models under the assumption that the neutrinos outside the
neutrino sphere flow freely after the apparent horizon formation. As
already mentioned, the time interval from the bounce to the apparent
horizon formation depends on the equation of state. In SYSC06, it is
shown that the features of the neutrino emission, such as a
neutronization burst, are not sensitive to the equation of state 
very much for the early phase. From Figure~\ref{lumi}, we can see
that the sign of neutronization burst becomes less remarkable and
disappears for the higher entropy models.

\begin{figure} 
\plotone{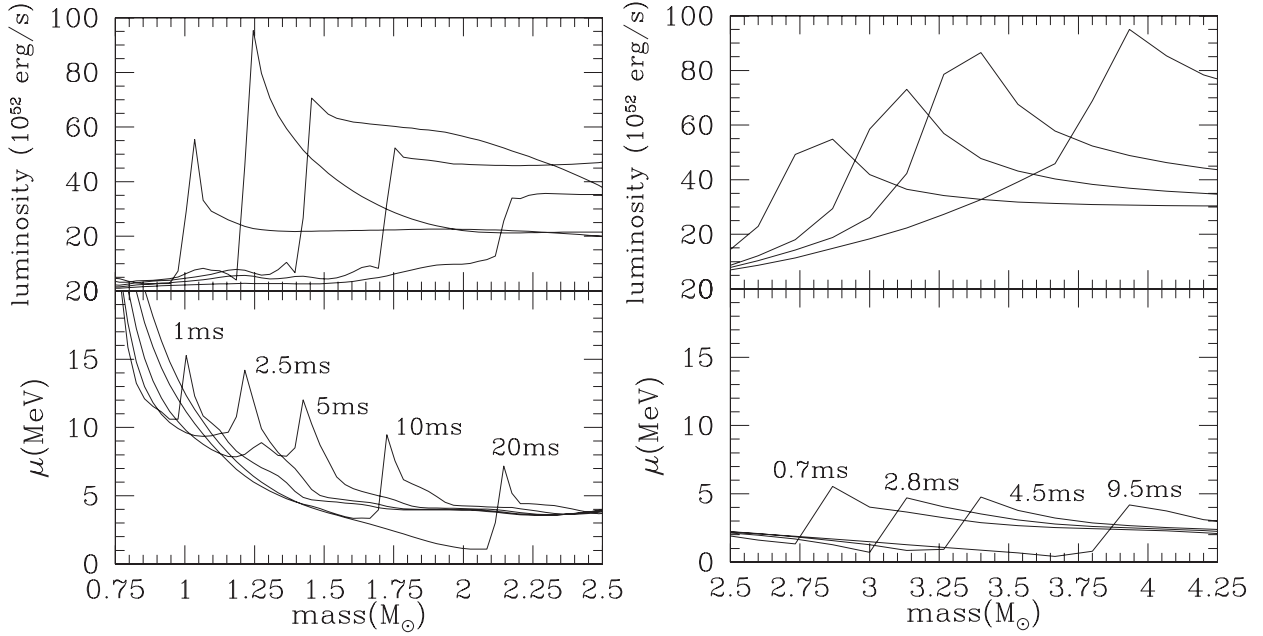} 
\caption{Snapshots of the profiles for the luminosity and the chemical
 potential of an electron-type neutrino. The left panel corresponds to
 the model~1a ($s=3k_\mathrm{B}$) and the right to the model~4a
 ($s=7.5k_\mathrm{B}$).}
\label{lumimu} 
\end{figure}

In order to analyze these features, we discuss the neutrino emission
from model~1a, as a reference model. In the upper left panel of
Figure~\ref{lumimu}, we show snapshots of the luminosity of an
electron-type neutrino as a function of the baryon mass coordinate. We
can recognize that neutrinos are emitted on the shock surface
mainly. The luminosity on the shock surface has a peak (e.g., at
$1.25M_{\odot}$ in the upper left panel of Figure~\ref{lumimu}), which
is similar to the situation for ordinary supernovae (e.g., Thompson et
al. 2003). In the following, we estimate the value of the luminosity
semi-analytically and compare it with the results of our numerical
simulations.

\begin{figure} 
\plotone{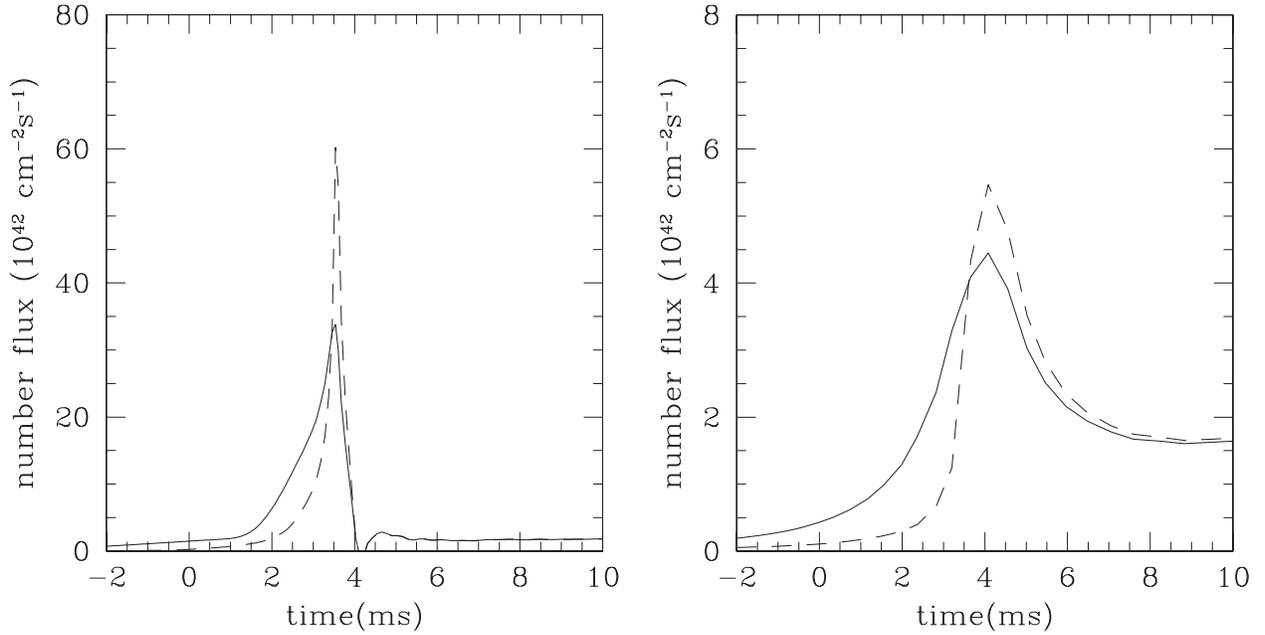} 
\caption{Time evolutions of the number flux for the electron-type
 neutrino with the energy $10~\mathrm{MeV}<E<20~\mathrm{MeV}$ detected
 by the comoving observer. Solid lines and dashed lines represent the
 results of our computation and the values estimated by the number
 density in equilibrium and $\langle\cos\theta\rangle$ of our
 computation, respectively. The left panel corresponds to model~1a
 ($s=3k_\mathrm{B}$) at $M=1.3M_\odot$ and the right to model~4a
 ($s=7.5k_\mathrm{B}$) at $M=3.3M_\odot$.}
\label{fluxcomp} 
\end{figure}

At first, the number density of neutrinos on the shock surface can be
evaluated roughly by the equilibrium value,
\begin{equation}
n_\mathrm{eq.}(\epsilon)d\epsilon\propto\frac{\epsilon^2}{\exp\left(\frac{\epsilon - \mu_\nu}{k_\mathrm{B}T}\right)+1}d\epsilon,
\label{neq}
\end{equation}
where $T$ and $\mu_\nu$ are the temperature and the chemical potential
of the electron-type neutrino in $\beta$-equilibrium at the shock
surface, respectively, and $k_\mathrm{B}$ is the Boltzmann
constant. Here the value $\mu_\nu$ is defined as
$\mu_\nu\equiv\mu_e-(\mu_n-\mu_p)$, where $\mu_e$, $\mu_n$ and $\mu_p$
are the chemical potentials of electron, neutron and proton, 
respectively, and they are given in the equation of state by Shen et
al. (1998a, 1998b). The number flux is estimated as
$cn_\mathrm{eq.}\langle\cos\theta\rangle$, where 
$\langle\cos\theta\rangle$ is a mean value of the angular cosine over
the neutrino angular distribution and $c$ is the light velocity. In
Figure~\ref{fluxcomp}, we compare the results of our numerical
computation with the number flux estimated above. We can see that the
equilibrium is not achieved completely, but the fraction is rather
constant, $\sim0.6$. Therefore, the luminosity is well estimated by
\begin{equation}
L(\epsilon)d\epsilon=C\frac{16\pi^2r^2\langle\cos\theta\rangle\epsilon^3}{h^3c^2\left(\exp\left(\frac{\epsilon - \mu_\nu}{k_\mathrm{B}T}\right)+1\right)}d\epsilon,
\label{lumies}
\end{equation}
where $h$ is the Planck constant, $r$ is the radius of the shock surface
and $C\sim0.6$.

From equation (\ref{lumies}), we can see that the luminosity is
determined by $r$, $\mu_\nu$, $T$ and
$\langle\cos\theta\rangle$. According to our numerical computation,
$T\sim1.5$~MeV and $\langle\cos\theta\rangle\sim0.5$ do not change very
much on the time scale of the neutronization burst. Thus, the luminosity
is dictated mainly by $r$ and $\mu_\nu$. Snapshots of the profiles of
$\mu_\nu$ are shown in the lower left panel of Figure~\ref{lumimu}, and
we can see that $\mu_\nu$ has a peak on the shock surface for the
following reason. When matter accretes onto the shock, the baryon mass
density and the electron number density rise, leading to the increase of
$\mu_e$ and, as a result, $\mu_\nu$. Immediately thereafter,
neutronization occurs and the value of $(\mu_n-\mu_p)$ rises, which
reduces $\mu_\nu$. We can recognize from Figure~\ref{lumimu} that the
peaks of $\mu_\nu$ and the luminosity are correlated. As for the time
evolution, the luminosity on the shock surface is lower at the early
phase because the shock radius is small. On the other hand, it is also
lower at the late phase because $\mu_\nu$ is lower. This is the reason
why the luminosity on the shock surface has a peak.

We now investigate model~4a, whose initial entropy is
$s=7.5k_\mathrm{B}$. In the lower right panel of Figure~\ref{lumimu},
snapshots of the profiles of $\mu_\nu$ for model~4a are shown. We can
see that the value of $\mu_\nu$ at the shock surface is lower than that
of model~1a at the early phase. This is because the baryon mass density
on the shock surface of model~4a at the bounce is lower than that of
model~1a, as has been mentioned. Accordingly, the electron number
density and $\mu_e$ are also lower for model~4a, and $\mu_\nu$ does not
rise so high. This is the main reason why the neutronization burst is
not remarkable. It is noted, moreover, that the electron fraction,
$Y_e$, on the shock surface of model~4a is lower than that of
model~1a. This is because nuclei do not exist and the nucleons are
already neutronized on the shock surface. The absence of nuclei is
consistent with the fact that the higher the initial entropy is, the
earlier nuclei dissolve into nucleons, as explained by
Figure~\ref{phase}. In addition, we can see that the luminosity of
$\nu_e$ rises monotonically. This is because the area of a shock surface
increases whereas $\mu_\nu$ is almost unchanging.

\begin{figure} 
\plotone{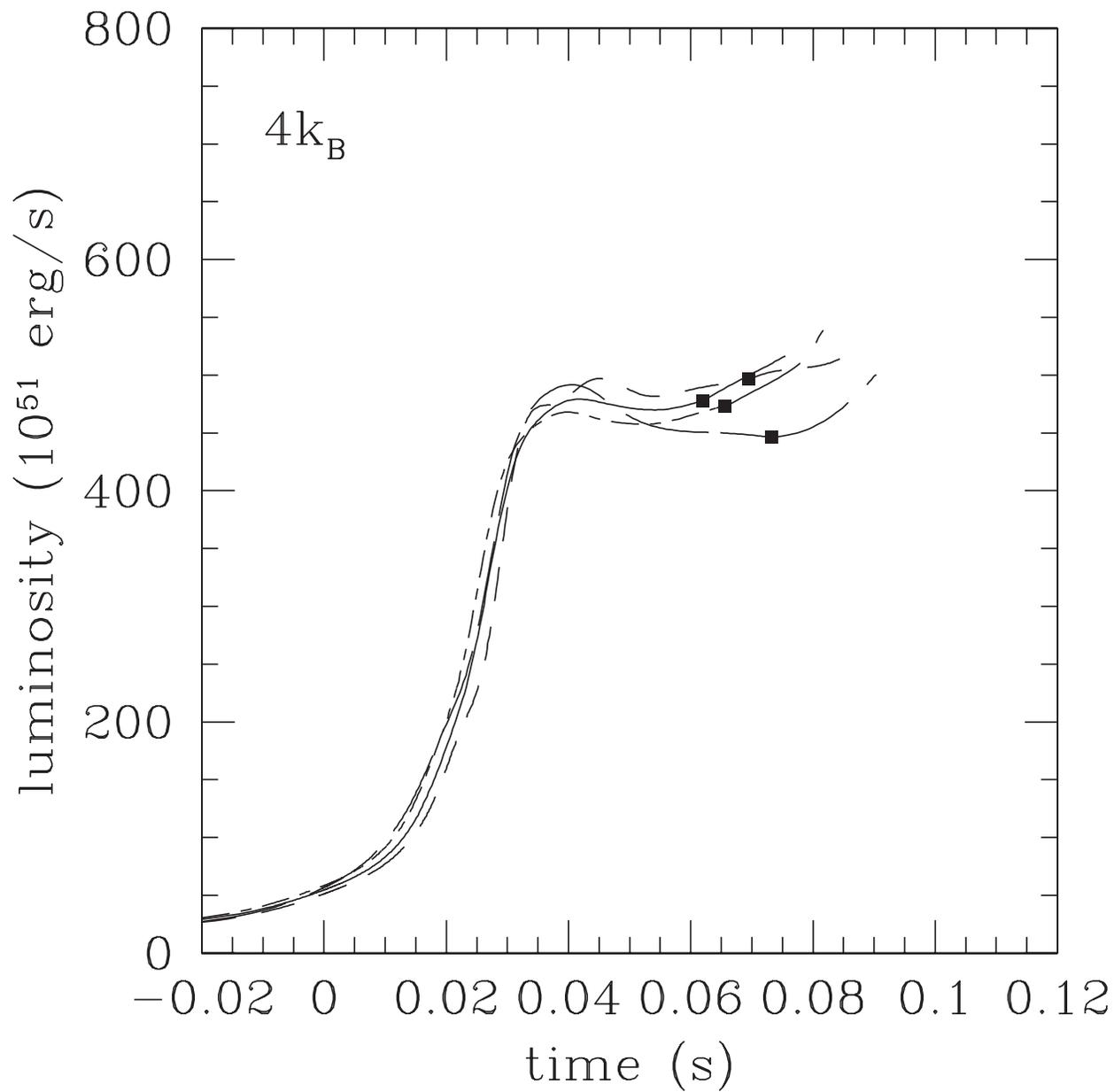} 
\caption{Luminosities of $\nu_e$ as a function of $t$ for
 models~2a (solid line), 2m (short-dashed line), 2e (long-dashed line)
 and 2g (dot-dashed line). The meaning of squares is the same as in
 Figure~\ref{lumi}.}
\label{far} 
\end{figure}

The results for the models with higher resolutions are shown in
Figure~\ref{far}. While the duration times of their neutrino emissions
differ slightly among the models as mentioned already, the profiles of
their neutronization bursts are not very different qualitatively. In
fact, the luminosity declines a little after the peak and increases
again for model~2a. This feature is well kept in other models with
higher resolutions.

\begin{figure} 
\plotone{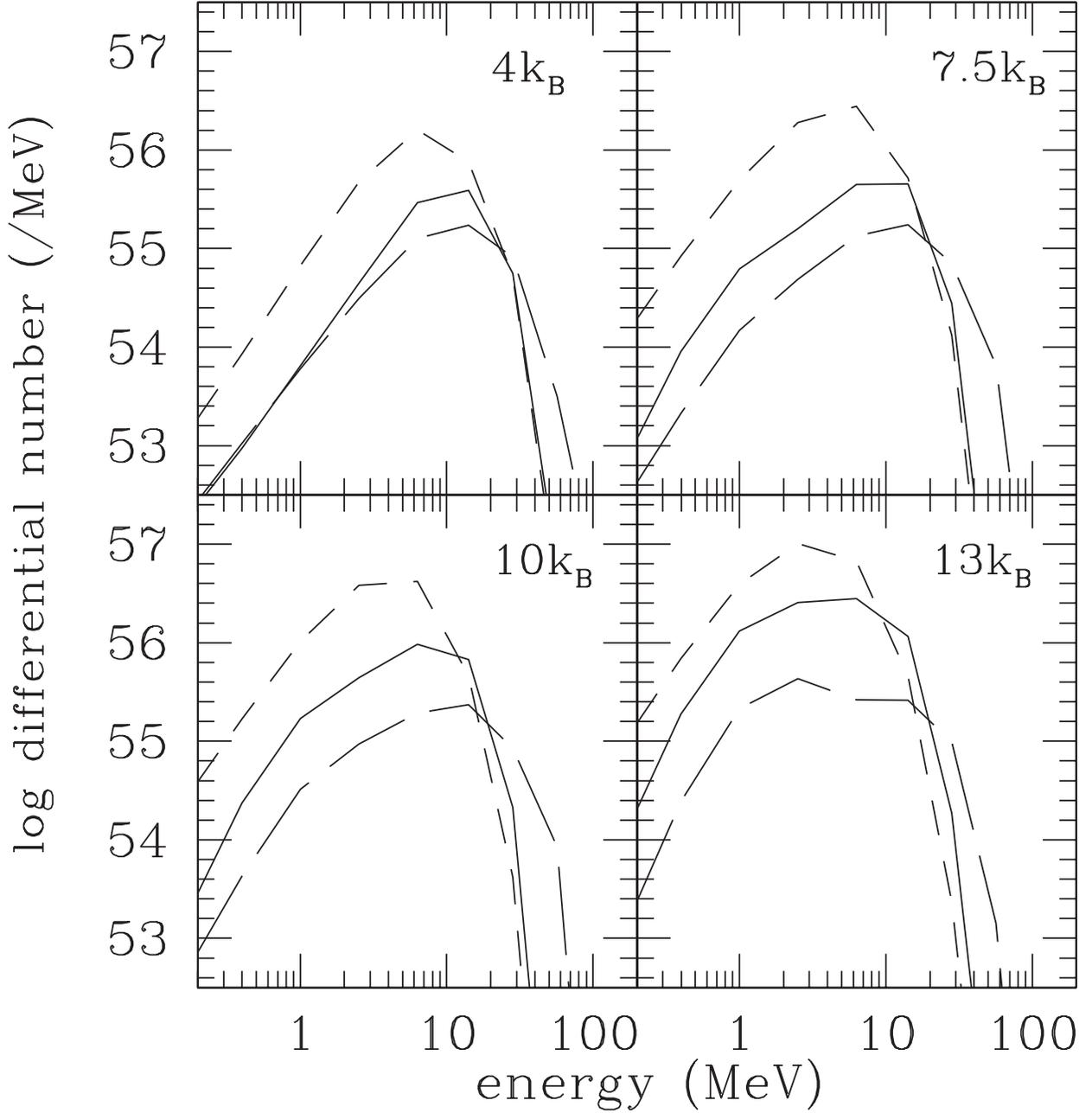} 
\caption{Spectra of time-integrated emissions of $\nu_e$ (short-dashed
 line), $\bar\nu_e$ (solid line) and $\nu_x$ (long-dashed line). Upper
 left, upper right, lower left and lower right panels are for models~2a
 ($s=4k_\mathrm{B}$), 4a ($s=7.5k_\mathrm{B}$), 5a ($s=10k_\mathrm{B}$)
 and 6a ($s=13k_\mathrm{B}$), respectively.}
\label{spect} 
\end{figure}

\begin{figure} 
\plotone{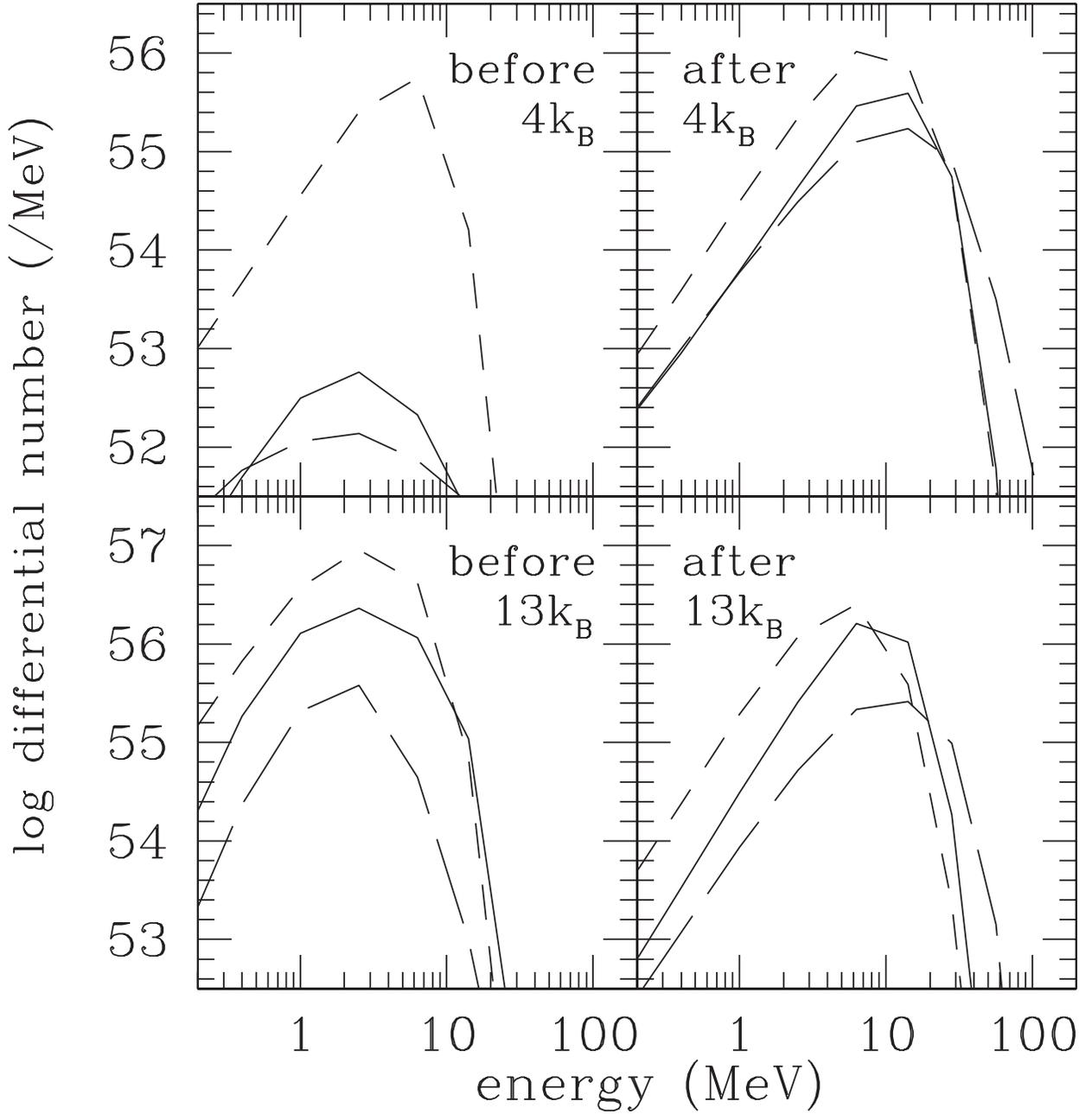} 
\caption{Spectra of time-integrated emissions of $\nu_e$ (short-dashed
 line), $\bar\nu_e$ (solid line) and $\nu_x$ (long-dashed line). The
 upper left and upper right panels give the time integrations of the
 emission before and after bounce, respectively, for model~2a
 ($s=4k_\mathrm{B}$). The lower left and lower right panels present the
 emission before and after shock formation, respectively, for model~6a
 ($s=13k_\mathrm{B}$).}
\label{spect2}
\end{figure}

We show the time-integrated neutrino spectra in Figure~\ref{spect}. We
can see that the spectra become softer for higher entropy models,
especially for $\bar\nu_e$ and $\nu_x$. In order to investigate this
tendency, we show the time-integrated spectra of the neutrino emitted
before and after the shock formation in Figure~\ref{spect2}. We can see
that, for higher entropy models, $\bar\nu_e$ and $\nu_x$ are also
emitted before the shock formation. They are created by the
electron-positron pair annihilation, and their energy is relatively
lower ($\lesssim$~several~MeV) because the temperature is low
($T\lesssim1$~MeV). On the other hand, for lower entropy models,
$\bar\nu_e$ and $\nu_x$ can not be produced by the electron-positron
pair process because positrons are absent owing to Pauli blocking. As
for the $\bar\nu_e$ and $\nu_x$ emitted after the shock formation, they
are mainly created by bremsstrahlung. In this phase, the temperature
near the neutrino sphere rises to $T\sim$~several~MeV, which makes the
neutrino energies relatively high: $\sim10$~MeV. Since the low energy
($\lesssim$~several~MeV) neutrinos are not emitted to any great extent
and the spectra become harder for lower entropy models, the emission of
low energy $\bar\nu_e$ and $\nu_x$ is characteristic for the collapse of
high entropy cores.

\subsection{Initial Velocity Dependence}\label{ivd}

\begin{figure} 
\plotone{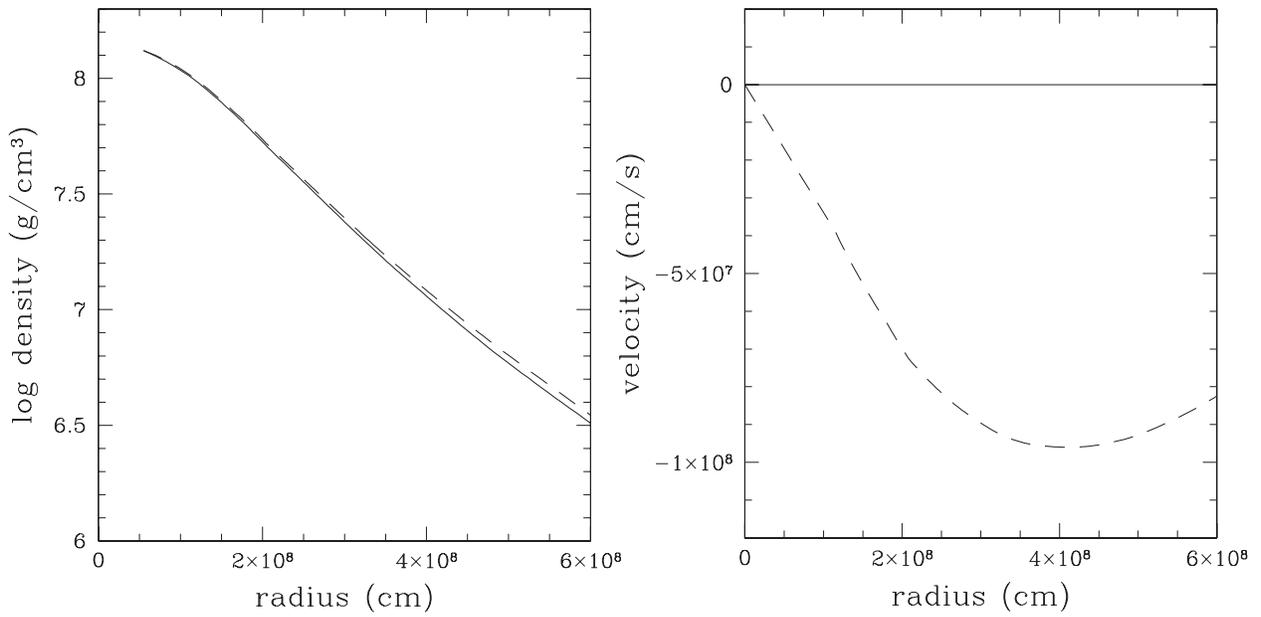} 
\caption{Comparisons of the density profiles (left) and the velocity
 profiles (right). Solid lines represent the initial profiles for
 model~2a; dashed lines represent the profiles for model~2b at the time 
 when the central density becomes the same as the initial central
 density of model~2a.}
\label{inipro} 
\end{figure}

We compare the results of models~2a and 2b, which are different in the
initial values of central density and temperature, but have the same
initial values of entropy per baryon ($s=4k_\mathrm{B}$). We can
consider that models~2a and 2b are the same model but with different
initial velocities, because the density profile of model~2b at the time
when the central density reaches that of the initial model of 2a almost
coincides with that of model~2a (Figure~\ref{inipro}). In reality, the
onset of a collapse is determined not only by the core structure but
also by the whole stellar structure. Thus, studying the initial velocity
dependence of the core is meaningful. 

As a result of this comparison, we find that the initial velocity does
not affect crucially the ensuing dynamics and the features of emitted
neutrinos such as total number spectra or the time evolutions of the
luminosity. This is because the velocity of model~2b at the time in
Figure~\ref{inipro} is several times lower than the sound speed at each
point. For instance, the fastest point of model~2b in
Figure~\ref{inipro} has the velocity
$\sim10^{8}\mathrm{cm\,\mathrm{s}^{-1}}$ while the sound speed is 
$\sim7\times10^{8}\mathrm{cm\,\mathrm{s}^{-1}}$, there. If the supersonic
region, where the infalling velocity exceeds the sound speed, existed in
the initial model, the initial velocity profile may be important for the
dynamics. However, since the temperature of our initial models is
slightly higher than the critical temperature for the
photodisintegration instability, they are unlikely to have supersonic
region.

\subsection{Collapse of Population III Star with $100M_{\odot}$}\label{rsm}

\begin{figure} 
\plotone{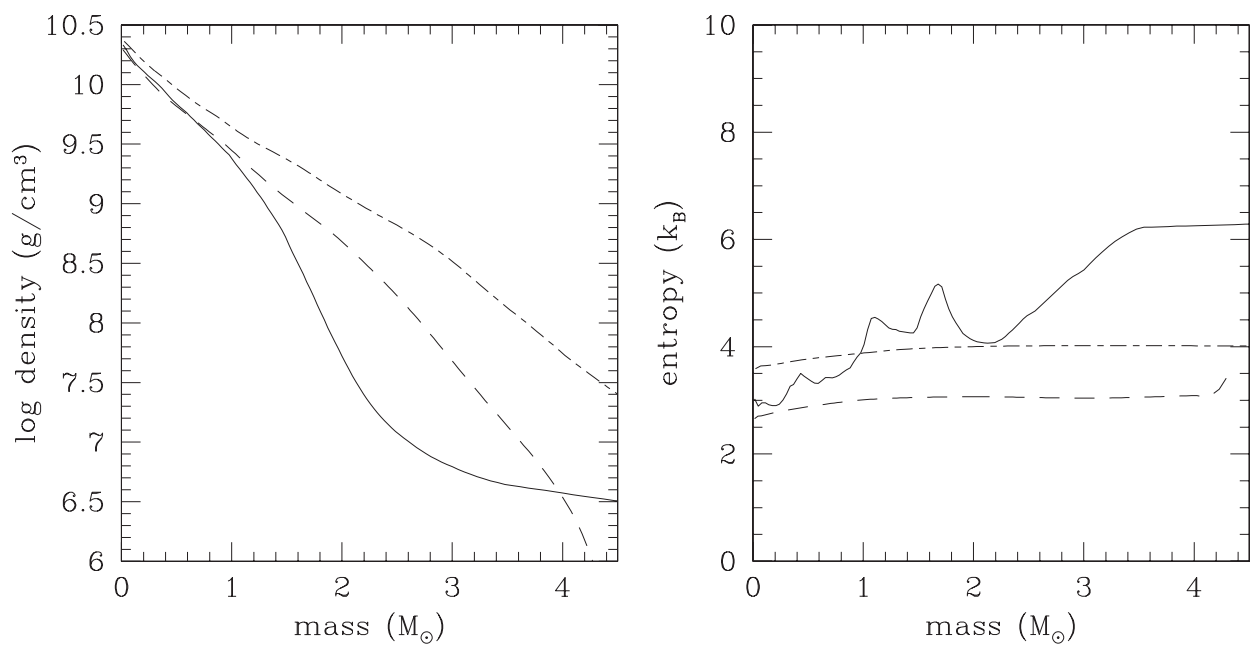} 
\caption{Comparisons of the density profiles (left) and the entropy
 profiles (right). Solid lines represent the initial profiles for model
 R and dashed and dot-dashed lines represent the profiles for models~1a
 and 2a, respectively, at the time when the central density becomes the
 same as the initial central density of model~R.}
\label{real} 
\end{figure}

In this section, we consider yet another example of very massive
stars, that is, a Population III star with $100M_{\odot}$. We use a
model constructed by Nomoto et al. (2005) with evolutionary
calculations, and we refer to it as model~R. This model is very massive
and its entropy at the center is higher than that of ordinary supernova
progenitors when it starts to collapse because the star does not lose
its mass at all in its evolution owing to its zero metallicity. It
should be emphasized that the isentropic models are meant for the
massive stars that may be produced in the present universe, for example,
by stellar mergers in clusters whereas Model~R corresponds to a
first-generation star in the past universe. Here we are interested in
the differences that these models may make. In Figure~\ref{real}, we
show the comparison of the initial state of model~R and our isentropic
models at the time when their central densities become the same as that
of model~R. We can recognize that model~R has the entropy
$\sim3.5k_\mathrm{B}$ in the central region, which is between those of
model~1a and 2a, whereas the iron core of model~R is smaller than that
of our models. In fact, the iron core mass of model~R is
$\sim2.32M_{\odot}$, which is close to that of model~1a. We show some of
the initial values at the center of model~R in
Table~\ref{bounce_result}. Incidentally, the initial velocity profile is
taken into account for model~R although it is much lower than the sound
speed at each point.

As a result of collapse, model~R has a bounce and recollapses to a black
hole. As shown in Table~\ref{bounce_result}, the values of the central
density and the central adiabatic index of model R at the bounce are
between those of model~1a and 2a. This suggests that these values are
determined by the initial central entropy as mentioned in
\S~\ref{bm}. On the other hand, model~R has a much longer time interval
from the bounce to the recollapse, compared with our models. This is
because the inner core mass of model~R at the bounce
($M_\mathrm{bounce}$) is smaller and the lower density of the outer core
(Figure~\ref{real}) gives lower accretion rates. For instance, at
$t=0.06$~s, model R has a mass accretion rate
$\sim4M_{\odot}\,\mathrm{s}^{-1}$ at the shock surface whereas model~1a
has $\sim11M_{\odot}\,\mathrm{s}^{-1}$. Thus it takes much time until
the inner core mass exceeds the maximum mass of the neutron star.

\begin{figure} 
\plotone{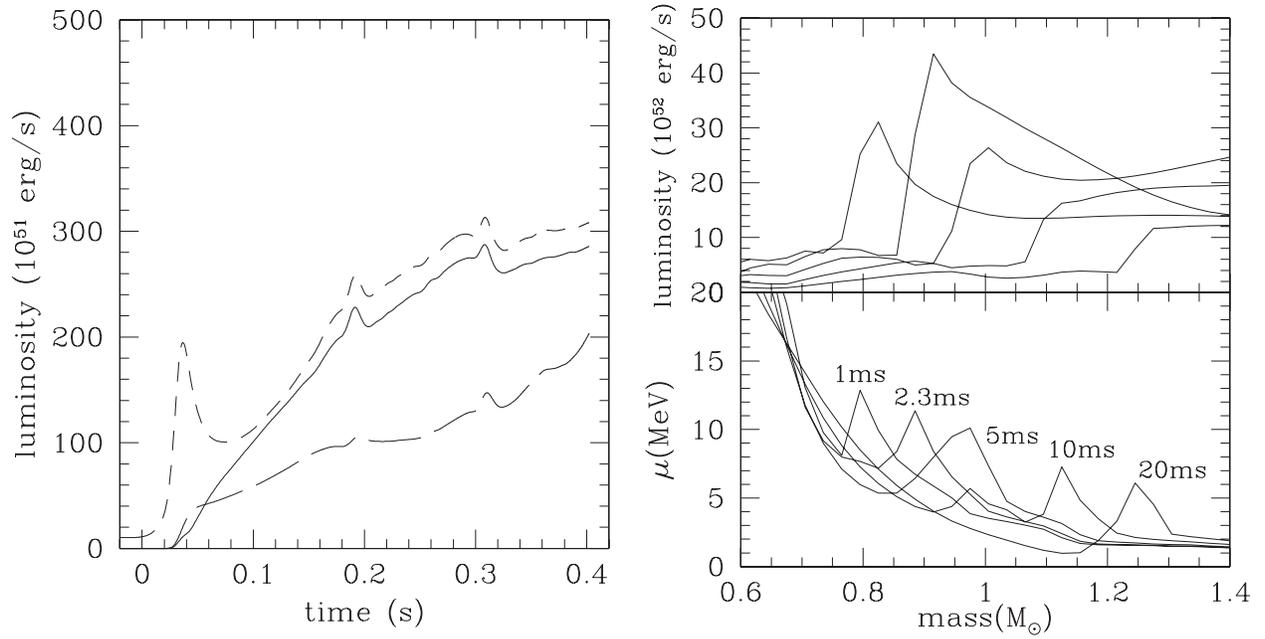} 
\caption{Results of the collapse for model R. In the left and right
 panel, the notations of lines are the same as in Figures~\ref{lumi}
 and \ref{lumimu}, respectively, but the end points of the lines in the
 left panel represent the time when the apparent horizon is formed.}
\label{real-lu} 
\end{figure}

We show the total energy of neutrinos emitted during the collapse of
model~R in Table~\ref{nutr-eg} and the time evolution of the emitted
neutrino luminosity in Figure~\ref{real-lu}. We can see that the total
energy of emitted neutrinos is larger than that of other isentropic
models, despite the fact that the neutrino luminosity of model~R is
lower than those of our models. This is because model~R neutrino
emission lasts much longer. Moreover, the mean energy of the emitted
neutrinos is larger for model~R. This is also due to the longer duration
time. The neutrino spectrum gets harder in the late phase because the
density of the accreting matter becomes lower and the temperature on the
neutrino sphere gets higher. Thus, the longer the duration time of
neutrino emission is, the larger the mean energy of the emitted
neutrinos becomes. It is noted that the duration time is sensitive to
the equation of state, which is already mentioned, and hence the total
and mean energy of emitted neutrinos is also sensitive to the equation
of state.

In the following, we discuss the features of the emitted neutrinos from
model~R for the early phase, which is not sensitive to the equation of
state as already mentioned. Comparing Figures~\ref{lumi} and
\ref{real-lu}, we can see that for model~R, the peak luminosity of the
electron-type neutrino by the neutronization burst is lower than those
of our models. The reason why it is lower than that of model 1a
($s=3k_\mathrm{B}$) is because the chemical potential of an
electron-type neutrino for model~R is lower than that for model~1a,
while the shock radii in both models are not so different from each
other (right panels of Figure~\ref{real-lu}). It is consistent with the
fact that the density at the bounce of model~R is lower than that of
model~1a (Table~\ref{bounce_result}). On the other hand, the shock radii
of models with $s\geq4k_\mathrm{B}$ (models~2a-4a) are larger than that
of model~R. This is the reason why the luminosity of the neutronization
burst for model~R is lower than those of models~2a-4a. Furthermore,
since the outer core density of model~R is much lower than those of
isentropic models (Figure~\ref{real}), $\mu_\nu$ drops quickly and the
shock radius does not get much larger after the neutronization. From
equation (\ref{lumies}), These features lead to the fact that the
luminosity of the electron-type neutrino after the neutronization burst
drops more steeply for model~R than for our models. It follows, then,
that the decline of the neutronization burst depends not only on the
initial entropy but also on the initial density profile. In particular,
the larger the initial density gradient is, the more steeply the
neutronization burst declines.

To sum up, the key parameters listed in Table~\ref{bounce_result} at
bounce (e.g. central density, temperature etc.) do not differ very much
between the isentropic models and model~R. This is not true for the time
profile of the neutronization burst because they depend not only on the
central density at bounce but also on the initial density
profile. However, since, in general, more massive iron cores have larger
entropies, the following trend is generally true: The neutronization
burst will become less remarkable as the progenitor gets more massive.

\subsection{Astrophysical Implications}\label{ai}

\subsubsection{Progenitor of IMBH}\label{prog}

For supermassive black holes (SMBH) located at the center of many
galaxies including ours, a new formation scenario via intermediate-mass
black holes (IMBH) has recently been suggested (e.g., Ebisuzaki et
al. 2001, Portegies Zwart et al. 2006). According to this scenario, very
dense stellar clusters are initially formed in the vicinity of the
galactic center ($\lesssim10$~pc), and the massive stars with
$\sim20M_\odot$ in them undergo runaway collisions to form IMBHs before
they lose most of their mass by supernova explosions and/or
pulsations. After that, these IMBHs merge together and finally form
SMBH. This scenario is supported by the discovery of the ultra luminous
X-ray compact sources in M82 galaxy, which indicate the existence of
IMBHs. It is conceivable that similar events occur in the Milky Way
Galaxy as mentioned later. This scenario assumes that the supermassive
stars formed by the runaway collisions would collapse to IMBHs when they
are $\sim1000M_\odot$.

Recently Suzuki et al. (2007) have studied the structures and evolutions
of these merged stars in the hydrogen burning. According to them, the
smaller star sits at the center of the larger star after the merger of
two stars with different masses. It is also demonstrated that the merged
stars become convectively unstable by the positive gradient of the mean
molecular weight and that their evolutions thereafter approach those of
the single homogeneous star with the same mass and abundance. The
central entropies of the merged stars will then be larger than those of
the inhomogeneous single stars with the same mass. This suggests the
possibility to form the IMBH progenitors by the merger without
experiencing the pair instability. Here we speculate the entropy of
these stars using previous studies on the single Population III
stars. Since the iron core of the Population III star with $100M_\odot$
has entropy of $\sim3.5k_\mathrm{B}$ (Nomoto et al. 2005), it is
expected that these IMBH progenitors have entropies
$\gtrsim3k_\mathrm{B}$.

Even if the pair instability occurs, the massive stars corresponding to
our models may still be formed. In fact, the positive entropy gradient
and/or rotation may suppress the convection in the merged star and the
entropy at the center may remain low after the merge. Then the merged
star has a massive envelope with a smaller core than the single stars
with the same total mass. If the pair instability occurs for these
objects, the nuclear burning may not produce total disruptions but lead
to the eventual collapse. Again inferring from single Population III
stars, we speculate that the central entropies of the merged stars will
be smaller than $s\sim16k_\mathrm{B}$, which corresponds to $300M_\odot$
in NSY06. It is incidentally mentioned that the relations between the
total mass and the iron core mass of merged IMBH progenitors is highly
uncertain at present.

In the preceding sections, we have shown that the neutrino signals from
the black hole formation are sensitive to the inner region of the
progenitor. In this section, assuming our models correspond to
above-mentioned merged stars which collapse to IMBHs at the center of
our Galaxy ($\sim8.5$~kpc from the sun), we estimate the neutrino event
number for the currently operating detectors.

As for the event rate of the IMBH formation, based on the
above-mentioned scenario and the fact that the SMBH residing in the
center of our Galaxy (SgrA$^\ast$) is $\sim3.5\times10^6M_\odot$ and the
age of our Galaxy is $\sim10$~Gyr, a very rough estimation for the
formation rate of IMBH with $\sim1000M_\odot$ is $\lesssim$~once per
1~Myr. It is, however, mentioned that this event rate may be
underestimated because star formation may not be continuous but
triggered by some environmental effects (e.g., the merger of
galaxies). Recent observations by Paumard~et al. (2006) have revealed
the existence of about 80 young massive stars within a distance of a
parsec from SgrA$^\ast$ and some of them are identified as OB stars and
their ages are about $6\pm2$ Myr. These facts indicate that stars are
actively formed in this region at present. Moreover, the IMBH candidate
with $\sim1300M_\odot$, IRS~13, is found in the same region (Maillard et
al. 2004). Thus, SgrA$^\ast$ may be currently growing under this
scenario.

In the following estimations for the neutrino event number, we do not
take into account the neutrino mixing, although it should be. Since the
mixing occurs mainly in the resonance regions and they are located
outside the iron core of the progenitor, the neutrino oscillation does
not affect the dynamics of core. Unfortunately the structures of the
envelopes of merged stars, which are crucial for the neutrino mixing,
are quite uncertain. There remain uncertainties as well on the mixing
parameters, such as the mixing angle of $\sin^22\theta_{13}$ or the mass
hierarchy. Thus, the precise evaluation of the neutrino flux including
the neutrino mixing is deferred to future study.

\subsubsection{Detection of low energy $\bar\nu_e$ by Super
   Kamiokande and KamLAND}

As already mentioned, a good deal of low energy $\bar\nu_e$ is emitted
from the collapse of the high entropy cores, which softens the
spectrum. We estimate the $\bar\nu_e$ event number for Super Kamiokande
III and KamLAND, currently operating neutrino detectors, under the
assumption that the black hole formations considered in former sections
occur at the center of our Galaxy. For both detectors, the dominant
reaction is the inverse beta decay,
\begin{equation}
\bar\nu_e + p  \longrightarrow  e^+ + n,
\label{ibd}
\end{equation}
which we take into account only. We adopt the cross section for this
reaction from Vogel \& Beacom (1999). For Super Kamiokande III, we
assume that the fiducial volume is 22.5 kton and the trigger efficiency
is 100\% at 4.5 MeV and 0\% at 2.9 MeV, which are the values at the end
of Super Kamiokande I (Hosaka et al. 2006). For KamLAND, we assume 1
kton fiducial mass, which means that $8.48\times10^{31}$ free protons
are contained (Eguchi et al. 2003). We also assume that the trigger
efficiency is 100\% for all $\bar\nu_e$ energy larger than the threshold
energy of the reaction.

The results are given in Table~\ref{antinue}. The total event number
does not change monotonically with the initial entropy of the core
because the total number of neutrinos depends on both the core mass and
the duration time of neutrino emission, as already mentioned. In order
to investigate the hardness of $\bar\nu_e$ spectrum, we calculate the
ratio of the event number by $\bar\nu_e$ with $<10$~MeV to that for all
events. The ambiguity about the distance of source is also canceled by
this normalization. This ratio gets larger as the entropy of the core
becomes higher. This suggests that we can probe the entropy of the black
hole progenitor especially in higher regimes ($s\geq7.5k_\mathrm{B}$)
because the event numbers of $\bar\nu_e$ with $<10$~MeV are over 100 by
Super Kamiokande III.

\subsubsection{Detection of neutronization burst by SNO}

The SNO detector consists of 1 kton of pure heavy water (D$_2$O) and can
distinguish $\nu_e$ flux by the charged-current reaction of the
deuterium disintegration. Since SNO can also detect the $\bar\nu_e$
flux, we can estimate the intensity of the neutronization burst by
comparing the event from the charged-current reaction of $\nu_e$,
\begin{equation}
\nu_e + d  \longrightarrow  p + p + e^-,
\label{cc1}
\end{equation}
and that of $\bar\nu_e$,
\begin{equation}
\bar\nu_e + d  \longrightarrow  n + n + e^+,
\label{cc2}
\end{equation}
using the SNO detector. SNO can also detect the neutral-current
reaction, 
\begin{equation}
\nu + d  \longrightarrow  n + p + \nu,
\label{nc}
\end{equation}
for all species. It is noted that the neutral-current reaction contains
$\nu_x$ ($=\nu_\mu$, $\bar\nu_\mu$, $\nu_\tau$ and $\bar\nu_\tau$) and
the neutrino sphere of $\nu_x$ differs more from that of $\nu_e$ than
that of $\bar\nu_e$ in general. Thus, for the comparison with reaction
(\ref{cc1}), reaction (\ref{cc2}) is more appropriate than reaction
(\ref{nc}). On the other hand, we also use (\ref{nc}) for the comparison
because the event number of (\ref{nc}) is larger than that of
(\ref{cc2}). In our calculation, we use the cross sections from Ying et
al. (1989) and assume that the trigger efficiency of these reactions is
100\%. In fact, it is $\sim$ 92\% these days, which is the neutron (in
the right hand side of equation (\ref{nc})) capture efficiency on
$^{35}$Cl and deuterons (Oser 2005). 

In the following analysis, we regard the emission of neutrinos before
$t=0.06$~s as the neutronization burst, where the time $t$ is measured
from the bounce. The criterion $t=0.06$~s is chosen empirically from our
simulations as an expedient. The method for extracting the
neutronization burst from detection should be reconsidered for more
detailed studies. Here we calculate the event numbers for $t<0.06$~s as
well as those for the entire duration time of the neutrino emission, and
the results are summarized in Table~\ref{bstnu}. We can recognize that
the ratios of the $\nu_e$ event number ($N_{\nu_e,<0.06\,\mathrm{s}}$)
to the total event number of the charged-current reactions
($N_{\nu_e,<0.06\,\mathrm{s}}+N_{\bar\nu_e,<0.06\,\mathrm{s}}$)
and that for the neutral-current reaction ($N_{NC,<0.06\,\mathrm{s}}$)
are larger for the models whose neutronization burst declines more
steeply. Despite the fact that these neutronization burst numbers are of
the order of 10, we can probe into the black hole progenitors in
principle.

It is finally noted that the estimations in the current study are based
on the spherically symmetric models. If progenitors are rotating
rapidly, the neutrino sphere will become non-spherical and the neutrino
emissions will be affected in general. This will be the subject of
future investigations.

\section{Conclusions}

In this paper, we have numerically studied gravitational collapse and
black hole formation of massive iron cores systematically, taking into
account the reactions and transports of neutrinos in detail. Massive
iron cores with $\sim10M_\odot$ have a bounce owing to thermal nucleons,
following which they collapse to black holes when the maximum mass is
reached. As for the emitted neutrinos, the spectra of $\bar\nu_e$ and
$\nu_x$ ($=\nu_\mu$, $\bar\nu_\mu$, $\nu_\tau$ and $\bar\nu_\tau$)
become softer for more massive models, or higher entropy models, because
a high entropy generates a large number of electron-positron pairs,
which create $\bar\nu_e$ and $\nu_x$. The neutronization burst from more
massive iron core becomes less remarkable or disappears completely. This
is because the density at the bounce is lower and even the $\nu_e$
number density in equilibrium becomes lower.

We have found that if the initial velocity is lower than the sound
speed, it does not affect the collapse very much. We have also compared
the collapse of our isentropic models with that of the realistic model,
which is obtained by the detailed modeling of the evolution of
Population III stars and we have found that the steep decline of the
neutronization burst depends not only on the initial entropy but also on
the initial density profile. Moreover, assuming our models as the
progenitors of IMBHs collapsing at the Galactic center, we have
estimated the neutrino event numbers. As a result, for Super Kamiokande
III, the ratio of the $\bar\nu_e$ event number for $<10$~MeV to that for
all events gets larger as the entropy of the core becomes higher,
especially for $s\geq7.5k_\mathrm{B}$. We have suggested that we can use
these features to probe into the progenitors. As for the lower entropy
cores, despite the fact that the event number for the early phase of the
emission is less than 100 by SNO, we have suggested that the steep
decline of the neutronization burst can be distinguished in principle.

Concerning the prediction of neutrino event number, there is a room
for further improvement. Firstly, the effects of the neutrino
oscillation should be taken into account. Secondly, multi-dimensional 
effects, such as rotation or magnetic field may be important, since they
will affect the dynamics of collapse itself. This study will be
hopefully prove a first step toward a neutrino astrophysics for black
holes.





\acknowledgments

We are grateful to Tadao Mitsui for valuable comments on the KamLAND
detector and Hideyuki Umeda for providing a realistic progenitor
model. We would like to thank Hideyuki Suzuki for fruitful
discussions. In this work, numerical computations were partially
performed on the Fujitsu VPP5000 at the Center for Computational
Astrophysics (CfCA) of the National Astronomical Observatory of Japan
(VPP5000 System projects wkn10b, ikn18b, iks13a), and on the
supercomputers in JAERI, YITP and KEK (KEK Supercomputer project 
108). This work was partially supported by Japan Society for Promotion
of Science (JSPS) Research Fellowship, Grants-in-Aid for the Scientific
Research from the Ministry of Education, Science and Culture of Japan
through 14740166, 15540243, 15740160, 17540267, 18540291, 18540295 and
the 21st-Century COE Program ``Holistic Research and Education Center
for Physics of Self-organization Systems.''

\clearpage

\begin{table}
\caption{Key Parameters for all Models.} 
\begin{center}
\begin{tabular}{lrrrrrrrrr}
\tableline\tableline
\multicolumn{1}{c}{}  & $s_\mathrm{initial}$ & $M_\mathrm{iron}$
 & $\rho_\mathrm{initial}$ &  $T_\mathrm{initial}$ & $M_\mathrm{bounce}$
 & $\rho_\mathrm{bounce}$ & $T_\mathrm{bounce}$ & $\gamma_\mathrm{bounce}$
 & $t_\mathrm{recollapse}$ \\
  model & ($k_\mathrm{B}$) & ($M_\odot$) & ($\mathrm{g\,cm^{-3}}$) & (K) & ($M_\odot$) & ($\mathrm{g\,cm^{-3}}$) & (MeV) &  & (msec) \\ \hline 
  1a & 3.0  &  2.44 &  2.71$\times10^{8}$ & 7.75$\times10^{9}$ &  0.75 &  1.95$\times10^{14}$ & 25.9 & 2.38 & 96.7 \\
  2a & 4.0  &  3.49 &  1.40$\times10^{8}$ & 7.75$\times10^{9}$ &  1.10 &  9.58$\times10^{13}$ & 26.9 & 1.89 & 62.0 \\
  2b & 4.0  &  2.93 &  7.00$\times10^{7}$ & 6.86$\times10^{9}$ &  1.05 &  9.90$\times10^{13}$ & 26.7 & 1.91 & 63.4 \\
  3a & 5.0  &  4.97 &  8.82$\times10^{7}$ & 7.75$\times10^{9}$ &   1.5 &  2.97$\times10^{13}$ & 19.4 & 1.58 & 52.6 \\
  4a & 7.5  &  10.6 &  4.20$\times10^{7}$ & 7.75$\times10^{9}$ &   2.7 &  3.00$\times10^{12}$ & 12.0 & 1.54 & 37.9 \\
  5a & 10.0 &  19.3 &  2.67$\times10^{7}$ & 7.75$\times10^{9}$ &  ---  &  ---  &  ---  &  ---  &  ---  \\
  6a & 13.0 &  34.0 &  1.84$\times10^{7}$ & 7.75$\times10^{9}$ &  ---  &  ---  &  ---  &  ---  &  ---  \\
  R  & 3.5  &  2.32 &  2.34$\times10^{10}$& 1.61$\times10^{10}$&  0.65 &  1.37$\times10^{14}$ & 22.7 & 2.19 &  402 \\ \hline
  2a & 4.0  &  3.49 &  1.40$\times10^{8}$ & 7.75$\times10^{9}$ &  1.10 &  9.58$\times10^{13}$ & 26.9 & 1.89 & 62.0 \\
  2m & 4.0  &  3.49 &  1.40$\times10^{8}$ & 7.75$\times10^{9}$ &  1.04 &  9.64$\times10^{13}$ & 27.5 & 1.88 & 69.5 \\
  2g & 4.0  &  3.49 &  1.40$\times10^{8}$ & 7.75$\times10^{9}$ &  1.05 &  9.66$\times10^{13}$ & 27.2 & 1.88 & 65.7 \\
  2e & 4.0  &  3.49 &  1.40$\times10^{8}$ & 7.75$\times10^{9}$ &  1.10 &  8.25$\times10^{13}$ & 25.9 & 1.80 & 73.2 \\ \hline
\end{tabular}
\end{center}
\label{bounce_result}
\tablecomments{$s_\mathrm{initial}$ is the initial value of the entropy
 par baryon. $M_\mathrm{iron}$ and $M_\mathrm{bounce}$ are the mass
 of initial iron core and inner core at the bounce ($t=0$),
 respectively. $\rho_\mathrm{initial}$ and $\rho_\mathrm{bounce}$ are the
 central density of the initial model and at the bounce,
 respectively. $T_\mathrm{initial}$ and $T_\mathrm{bounce}$ are the
 central temperature of the initial model and at the bounce,
 respectively. $\gamma_\mathrm{bounce}$ is the central adiabatic index
 at the bounce. $t_\mathrm{recollapse}$ is the interval time from the
 bounce to the apparent horizon formation.} 
\end{table}

\begin{deluxetable}{lccccccc}
\tabletypesize{\scriptsize}
\tablewidth{0pt}
\tablecaption{Estimates of Average and Total Energies of Emitted Neutrinos.}
\tablehead{ & $\langle{E_{\nu_e}\rangle}$ &
 $\langle{E_{\bar \nu_e}\rangle}$ & $\langle{E_{\nu_x}\rangle}$ & 
 $E^\mathrm{tot}_{\nu_e,52}$ & $E^\mathrm{tot}_{\bar \nu_e,52}$ &
 $E^\mathrm{tot}_{\nu_x,52}$ & $E^\mathrm{tot}_{\mathrm{all},52}$ \\
 model & (MeV) & (MeV) & (MeV) & ($10^{52}\mathrm{ergs}$) & ($10^{52}\mathrm{ergs}$) & ($10^{52}\mathrm{ergs}$)  & ($10^{52}\mathrm{ergs}$) }
\startdata
 1a & 11.01 - 11.01 & 15.03 - 15.03 & 19.60 - 19.60 & 3.29 - 3.29 & 1.94 - 1.94 &  1.43 - 1.43  & 10.96 - 10.96  \\
 2a & 10.32 - 10.32 & 14.29 - 14.30 & 19.30 - 19.30 & 3.21 - 3.21 & 1.58 - 1.58 &  1.33 - 1.33 &  10.12 - 10.12 \\
 2b & 10.17 - 10.17 & 14.27 - 14.27 & 19.35 - 19.35 & 3.24 - 3.24 & 1.63 - 1.63 &  1.36 - 1.36 &  10.29 - 10.29 \\
 3a &  9.29 - 9.29 & 13.79 - 13.79 &  19.31 - 19.31 & 3.10 - 3.10 & 1.48 - 1.48 &  1.30 -  1.30 & 9.78 - 9.78 \\ 
 4a &  7.30 - 7.30 & 11.95 - 11.95 &  19.55 - 19.55 & 3.19 - 3.19 & 1.56 - 1.56 &  1.35 -  1.36 & 10.18 - 10.21 \\
 5a &  6.24 - 6.25 & 10.34 - 10.37 &  18.37 - 18.70 & 4.01 - 4.02 & 2.34 - 2.35 &  1.69 - 1.73  & 13.11 - 13.31 \\
 6a &  5.24 - 5.25 & 8.14 - 8.19 &  14.33 - 14.33 & 6.15 - 6.17 & 4.71 - 4.75 &  1.70 - 1.73 & 17.66 - 17.84 \\ 
 R  & 15.34 - 15.34 & 18.90 - 18.90 &  23.42 - 23.42 & 9.42 - 9.42 & 7.89 - 7.89 &  4.40 - 4.40 & 34.89 - 34.90 \\ 
\enddata
\label{nutr-eg}
\tablecomments{The mean energy of emitted $\nu_i$ (with upper and lower
 limits) is denoted as $\langle{E_{\nu_i}\rangle} \equiv
 E^\mathrm{tot}_{\nu_i}/N^\mathrm{tot}_{\nu_i}$, where
 $E^\mathrm{tot}_{\nu_i}$ and $N^\mathrm{tot}_{\nu_i}$ are the total
 energy and number of neutrinos. $E^\mathrm{tot}_{\mathrm{all}}$ is the
 total energy summed over all species.}
\end{deluxetable}

\begin{table}
\caption{Event Numbers of $\bar\nu_e$ from Super Kamiokande III and KamLAND.} 
\begin{center}
\begin{tabular}{lrrrr}
\tableline\tableline
\multicolumn{1}{c}{model}  
 & $\frac{N_{\bar \nu_e<\mathrm{10\,MeV,SK}}}{N_{\bar \nu_e,\mathrm{SK}}}$
 & $N_{\bar \nu_e,\mathrm{SK}}$
 & $\frac{N_{\bar \nu_e<\mathrm{10\,MeV,Kam}}}{N_{\bar \nu_e,\mathrm{Kam}}}$
 & $N_{\bar \nu_e,\mathrm{Kam}}$ \\ \hline 
  1a & 3.3\%  &  6163 &  3.3\%  & 174 \\
  2a & 4.0\%  &  4778 &  4.0\%  & 135 \\
  2b & 4.0\%  &  4910 &  4.0\%  & 139 \\
  3a & 4.6\%  &  4319 &  4.6\%  & 122 \\
  4a & 7.3\%  &  4018 &  7.3\%  & 114 \\
  5a & 11.8\% &  5326 &  12.0\% & 151 \\
  6a & 20.1\% &  9139 &  20.5\% & 259 \\ \hline
\end{tabular}
\end{center}
\label{antinue}
\tablecomments{The subscript``$<10$~MeV'' means the event of $\bar\nu_e$
 with $<10$~MeV, and the subscript ``SK'' and ``Kam'' mean the
 prediction for Super Kamiokande III and KamLAND, respectively.}
\end{table}

\begin{deluxetable}{lrrrrrrrr}
\tabletypesize{\scriptsize}
\tablewidth{0pt}
\tablecaption{Event Numbers by SNO.} 
\tablehead{ model
 & $N_{\nu_e,<0.06\,\mathrm{s}}$
 & $N_{\bar\nu_e,<0.06\,\mathrm{s}}$
 & $\frac{N_{\nu_e,<0.06\,\mathrm{s}}}{N_{\nu_e,<0.06\,\mathrm{s}}+N_{\bar\nu_e,<0.06\,\mathrm{s}}}$
 & $N_{\mathrm{NC,}<0.06\,\mathrm{s}}$
 & $\frac{N_{\mathrm{\nu_e,}<0.06\mathrm{s}}}{N_{\mathrm{NC,}<0.06\,\mathrm{s}}}$
 & $N_\mathrm{\nu_e,all}$ & $N_\mathrm{\bar\nu_e,all}$
 & $N_\mathrm{NC,all}$}
\startdata
  1a & 30.7 &  6.4 & 82.7\% & 45.4 & 67.5\% & 84.2 & 45.4 &  201 \\
  2a & 40.8 & 10.7 & 79.2\% & 70.4 & 57.9\% & 69.8 & 30.1 &  162 \\
  2b & 42.6 & 12.4 & 77.5\% & 77.1 & 55.3\% & 73.6 & 33.8 &  179 \\
  3a & 38.4 & 12.6 & 75.3\% & 78.5 & 49.0\% & 56.7 & 24.7 &  142 \\
  4a & 31.8 & 15.1 & 67.9\% & 91.2 & 34.9\% & 37.4 & 19.3 &  124 \\
  5a & --- & --- & --- & --- & --- & 44.9 & 32.7 & 220 \\
  6a & --- & --- & --- & --- & --- & 59.3 & 54.0 & 221 \\
\enddata
\label{bstnu}
\tablecomments{These values are the event number for the charged-current
 reaction except $N_{\mathrm{NC,}<0.06\mathrm{s}}$ and
 $N_\mathrm{NC,all}$. The subscript``$<0.06$~s''  means the event at
 $t<0.06$~s, where $t$ is the time measured from the bounce, and the
 subscript ``all'' means the event for all duration times of neutrino
 emission.}
\end{deluxetable}


\begin{thebibliography}{}
\bibitem[]{}
 Ebisuzaki, T., et al. 2001, \apj, 562, L19
\bibitem[]{}
 Eguchi, K., et al. 2003, Phys. Rev. Lett., 90, 021802
\bibitem[]{}
 Fryer, C. L. 1999, \apj, 522, 413
\bibitem[]{}
 Fryer, C. L., Woosley, S. E., \& Heger, A. 2001, \apj, 550, 372  
\bibitem[]{}
 Heger, A., Fryer, C. L., Woosley, S. E., Langer, N \& Hartmann, D. H. 2003, \apj, 591, 288  
\bibitem[]{}
 Herant, M., Benz, W., Hix, W. R., Fryer, C. L., \& Colgate, C. A. 1994, \apj, 435, 339 
\bibitem[]{}
 Hosaka, J., et al. 2006, Phys. Rev. D, 73, 112001
\bibitem[]{}
 Linke, F., Font, J. A., Janka, H.-Th., M$\ddot{\mathrm{u}}$ller, E., \& Papadopoulos, P. 2001, A\&A, 376, 568
\bibitem[]{}
 Maillard, J. P., Paumard, T., Stolovy, S. R., \& Riguaut, F. 2004, A\&A, 423, 155
\bibitem[]{}
 Misner, C. W., \& Sharp, D. H. 1964, Phys. Rev., 136, 571
\bibitem[]{}
 Nakamura, F., \& Umemura, M. 2001, \apj, 548, 19
\bibitem[]{}
 Nakazato, K., Sumiyoshi, K., \& Yamada, S. 2006, \apj, 645, 519
\bibitem[]{}
 Nomoto, K., Tominaga, N., Umeda, H., Maeda, K., Ohkubo, T., Deng, J., \& Mazzali, P. A. 2005, ASP Conf. Ser., 332, 374
\bibitem[]{}
 Ohkubo, T., Umeda, H., Maeda, K., Nomoto, K., Tsuruta, S., \& Rees, M. J. 2006, \apj, 645, 1352
\bibitem[]{}
 Oser, S. M. 2005, Nucl. Phys., A758, 677c
\bibitem[]{}
 Paumard, T., et al. 2006, \apj, 643, 1011
\bibitem[]{}
 Portegies Zwart, S. F., Makino, J., McMillian, S. L. W., \& Hut, P., 1999, A\&A, 348, 117
\bibitem[]{}
 Portegies Zwart, S. F., Baumgardt, H., McMillian, S. L. W., Makino, J., Hut, P., \& Ebisuzaki, T. 2006, \apj, 641, 319
\bibitem[]{}
 Sekiguchi, Y. I., \& Shibata, M. 2005, Phys. Rev. D, 71, 084013
\bibitem[]{}
 Shen, H., Toki, H., Oyamatsu, K., \& Sumiyoshi, K. 1998a, Nucl. Phys., A637, 435
\bibitem[]{}
 Shen, H., Toki, H., Oyamatsu, K., \& Sumiyoshi, K. 1998b, Prog. Theor. Phys., 100, 1013
\bibitem[]{}
 Sumiyoshi, K., Yamada, S., Suzuki, H., Shen, H., Chiba, S., \& Toki, H. 2005, \apj, 629, 922
\bibitem[]{}
 Sumiyoshi, K., Yamada, S., Suzuki, H., \& Chiba, S., 2006, Phys. Rev. Lett., 97, 091101
\bibitem[]{}
 Suzuki, T. K., Nakasato, N., Baumgardt, H., Ibukiyama, A., Makino, J., \& Ebisuzaki, T. 2007, astro-ph/0703290, submitted to \apj
\bibitem[]{}
 Thompson, T. A., Burrows, A., \& Pinto, P. A. 2003, \apj, 592, 434 
\bibitem[]{}
 van Riper K. A. 1979, \apj, 232, 558
\bibitem[]{}
 Vogel, P., \& Beacom, J. F. 1999, Phys. Rev. D, 60, 053003
\bibitem[]{}
 Yamada, S. 1997, \apj, 475, 720
\bibitem[]{}
 Yamada, S., Janka, H.-Th., \& Suzuki, H. 1999, A\&A, 344, 533
\bibitem[]{}
 Ying, S., Haxton, W. C., \& Henley, E. M. 1989, Phys. Rev. D, 40, 3211
\end{thebibliography}
\end{document}